\documentclass[fleqn,usenatbib]{mnras}

\usepackage{newtxtext,newtxmath}

\usepackage[T1]{fontenc}
\usepackage{ae,aecompl}

\usepackage{graphicx}	
\usepackage{amsmath}	
\usepackage{amssymb}	

\usepackage{tabularx}   
\usepackage{adjustbox}  
\usepackage{pdflscape}
\usepackage[graphicx]{realboxes}
\usepackage{rotating}
\usepackage{soul}       

\graphicspath{{./}{}}

\title[Abundance analyses of Li-enriched and normal giants in the {\it GALAH} survey] {Abundance analyses of Li-enriched and normal giants in the {\it GALAH} survey}

\author[Deepak, Lambert \& Reddy]
{Deepak,$^{1,2}$\thanks{E-mail: deepak@iiap.res.in, deepak4astro@gmail.com}
David L. Lambert,$^{3}$\thanks{E-mail: dll@astro.as.utexas.edu}
Bacham E. Reddy$^{1}$\\
$^1$Indian Institute of Astrophysics, Bangalore -- 560034, India\\
$^2$Pondicherry University, R. V. Nagara, Kalapet, Puducherry -- 605014, India\\
$^3$W.J. McDonald Observatory and Department of Astronomy, The University of Texas at Austin, Austin, TX 78712, USA}
\date{Accepted 2020 March 09. 2020 March 09; in original form 2019 December 13}

\pubyear{2019}

\begin{document}
\label{firstpage}
\pagerange{\pageref{firstpage}--\pageref{lastpage}}
\maketitle

\begin{abstract}
Compositions of lithium-enriched and normal giants among the {\it GALAH} survey are compared. Except for Li, the only detectable abundance difference between lithium-enriched and normal giants among the investigated elements from carbon to europium occurs for carbon. Among Li-rich giants with A(Li) = 1.8 to 3.1, the C deficiency is very similar to that reported for the normal giants (with A(Li) $<$ 1.8) where the slight C deficiency arises from the first dredge-up. Carbon is slightly under abundant relative to normal giants among the super Li-rich giants where the Li abundance exceeds A(Li) = 3.2. The C abundance as well as the $^{12}$C/$^{13}$C ratios from the literature suggest that addition of Li to create a Li-rich giant may occur independently of the abundance changes wrought by the first dredge-up. Creation of a super Li-rich giant, however, appears to occur with additional CN-cycle conversion of C to N.
The probability of becoming a Li-rich giant is approximately independent of a star's mass, although the majority of the Li-rich giants are found to be low mass ($M \leq$ 2 M$_\odot$). The frequency of occurrence of Li-enriched giants among normal giants is about one percent and slightly dependent on metallicity ([Fe/H]).
Li-enriched and normal giants are found to have similar projected rotational velocity which suggest that Li-enrichment in giants is not linked to scenarios such as mergers and tidal interaction between binary stars.

\end{abstract}




\section{Introduction} \label{sec:introduction}  

Lithium provides a wonderful array of astrophysical problems for both theoreticians and observers interested in the physics of stellar interiors and atmospheres. One such problem concerns the presence of lithium-rich low-mass giants.  In a low mass main sequence star, lithium survives in the thin outer skin. Then this surviving lithium is greatly diluted as the convective envelope grows deeper on a star becoming a red giant. Given that the lithium abundance in many young stars is now about A(Li) = 3.2 dex\footnote{Elemental abundances are given on the traditional scale, e.g., A(X) = $\log$(N(X)/N(H)) $+$ 12 where N(X) is the number density of element.}
and a solar metallicity giant's convective envelope may induce a dilution by a factor of about 30 at 1 $M_\odot$ to 60 at 1.5 $M_\odot$ 
\citep{Iben1967}, the red giant may be expected to have a maximum surface abundance  A(Li)$\simeq1.8$ at 1 $M_\odot$ to $1.5$ at 1.5 $M_\odot$ where the realization is that many giants show  a lower abundance because many main sequence stars including the Sun have depleted their surface Li below the above initial value. In this paper, we discuss a rare class of low-mass giants which has a lithium abundance  in excess of the expected result from the first dredge-up, say A(Li) $>1.8$. These giants are known as Li-rich giants and extreme examples -- say, with A(Li) $>3.2$ -- are referred to as super Li-rich giants. 
Lithium enrichment is also seen in more luminous and cooler giants well advanced in evolution along the asymptotic giant branch (AGB) where the third dredge-up occurs. Early discovery of very strong Li 6707 \AA\ resonance lines in the carbon stars WZ Cas and WX Cyg \citep{Sanford1944ApJ....99..145S,Sanford1950ApJ...111..262S} preceded understanding of the AGB and its complexities enabling  conversion of an O-rich to a C-rich stellar atmosphere. Li enrichment in similarly luminous but O-rich AGB giants is now observed \citep{SmithLambert1989}. This paper does not consider such AGB stars but is directed at understanding Li-rich primarily K giants on the red giant branch (RGB) and early AGB.

Discovery of Li-rich K giants is traceable to \cite{WallersteinSneden1982}'s analysis of the K III giant HD 112127 where the Li abundance was found to be ``approximately the maximum seen in unevolved stars" (i.e., A(Li) $\sim3.2$) in a star clearly subjected to the first dredge-up, i.e., the $^{12}$C/$^{13}$C is well below its likely initial value. The rarity of Li-rich K giants was clearly established by the \cite{BrownSnedenLambert1989} survey of some 600 giants and a finding that only 9 were found to be Li-rich (i.e., A(Li) $\geq1.8$).
Subsequently, other discoveries of Li-rich K giants have been reported: see, the compilation of about 150 reported discoveries  by \cite{CaseyRuchtiMasseronRandich2016}. Presently, the most Li-rich K giant known is TYC 429-2097-1 with A(Li) $=4.51$ \citep{YanHL2018Nature}. The approximately 2000 Li-rich K giants ($\log$(Li) $\geq1.5$)  extracted from the huge {\it LAMOST} low-resolution survey \citep{Casey2019ApJ...880..125C} correspond to an occurrence rate of about 1\%.

A theoretical accounting for the Li abundance of Li-rich giants must also offer an explanation for Li abundances up to 1000-times greater than the Li left in the atmosphere after the first dredge-up. One imaginative solution with a contemporary flavour precedes discovery of Li-rich giants: the capture by the giant of terrestrial planetary material with its original Li abundance \citep{Alexander1967Obs....87..238A}. Quite different solutions call on the star's internal supply of $^3$He and its partial conversion in the interior via $^3$He($\alpha,\gamma)^7$Be with convection sweeping material out to lower temperatures allowing $^7$Be to undergo electron-capture to $^7$Li, the so-called Cameron-Fowler (\citeyear{CameronFowler1971})  mechanism. The $^3$He present in the star at its birth may be supplemented in low-mass stars by $^3$He synthesized in the main sequence phase through the initial stages of the H-burning $pp$-chain. Successful transformation of $^3$He in the stellar interior to $^7$Be and the appearance of $^7$Li at the stellar surface depends critically on the strength and duration of convection  in the envelope; both $^7$Be and $^7$Li will be destroyed by protons in the deeper reach of the convective envelope. Critical too is the trigger of the Cameron-Fowler mechanism which may lie within the star or be provided by presence of a companion star, as suggested by \cite{Casey2019ApJ...880..125C}. Invocation of the Cameron-Fowler mechanism must also explain why so few giants exhibit Li enrichment. Limiting possibilities may be: is the mechanism triggered only under special circumstances or does the synthesized Li survive but a short time?

\begin{figure}
\includegraphics[width=0.5\textwidth]{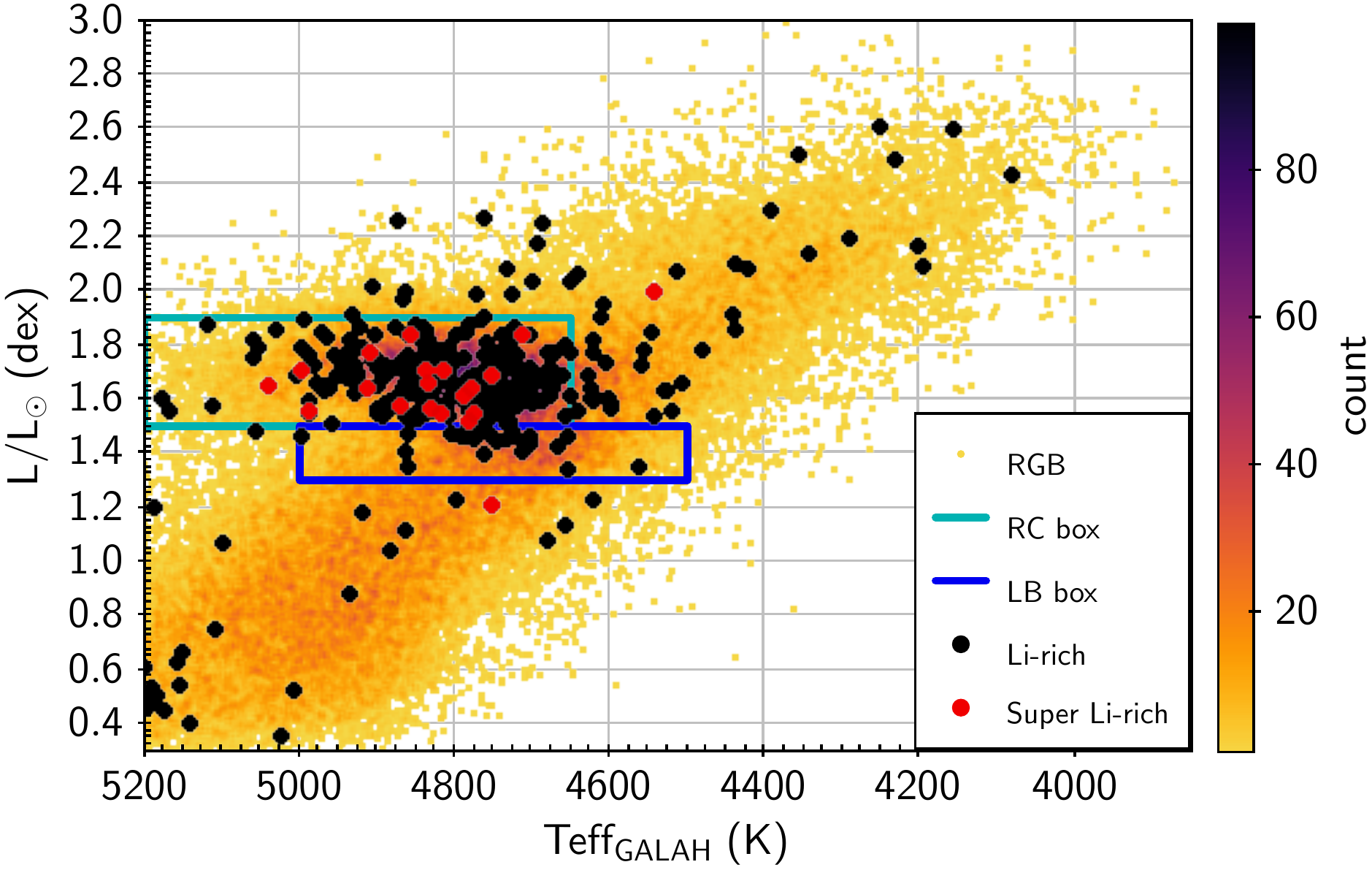}
\caption{HR diagram showing location of Li-rich giants (1.8 $\leq$ A(Li) $\leq$ 3.2 dex) (black dots) and super Li-rich giants (A(Li) $>$ 3.2 dex) (red dots), with the complete  sample of giants in the background.  The outlines of the two most prominent evolutionary regions --  the red-clump (RC) and the luminosity bump (LB) -- are taken from Deepak \& Reddy (2019).  Some of the cooler giants within the RC luminosity range are not included here to avoid  possible contamination between RC and LB.
\label{fig:f1}}
\end{figure}

A clue to origins of  lithium enrichment may be provided by the luminosity distribution of the Li-rich giants. A suspicion  held since the original survey by \cite{BrownSnedenLambert1989} and later by
\cite{KumarReddyLambert2011} and promoted repeatedly recently  has now been wonderfully confirmed by \cite{DeepakReddyBE2019MNRAS.484.2000D} using Li abundances provided by the {\it GALAH} survey,  by \cite{Casey2019ApJ...880..125C} by using the {\it LAMOST} survey, and by \cite{SinghReddyBharat2019ApJ...878L..21S} based on a survey of 12,500 giants that are common among the {\it LAMOST} spectroscopic and the {\it Kepler} time-resolved photometric survey catalogues. The latter study makes use of the robust astroseismic technique of separating giants of He-core burning from those of ascending the RGB with their inert He-core \citep{BeddingMosser2011Natur}.
The great majority of the Li-rich giants  are  relatively long-lived He-core burning (`clump') giants. A few have a luminosity above the clump, i.e., are on the early-AGB but prior to the onset of the third dredge-up when activation of the Cameron-Fowler mechanism is predicted and observed to provide strong Li enhancements  in very luminous giants \citep{SmithLambert1989}. In addition, a few Li-rich giants have the luminosity of the ``luminosity bump" on the red giant  branch well prior to the clump phase. 
The bump on the RGB occurs at the point when the H-burning shell moves outwards across the H-discontinuity. 
At this point in its evolution, models show that a star's evolution reverses course and moves down the RGB slightly before reversing again and continuing  to evolve up the RGB.
\footnote{The referee pointed out that the luminosity bump is also known as the `Thomas peak' \citep{Thomas1967ZA.....67..420T, Demarque1987ASSL..137..121D}.}
Li-rich giants at luminosities below that of clump and bump stars are assumed to have yet to complete their first dredge-up. 
Figure \ref{fig:f1} shows the H-R diagram for all giants with Li-rich (A(Li) $\geq$ 1.8 dex) and super Li-rich giants (A(Li) $\geq$ 3.2 dex) highlighted, drawn by \cite{DeepakReddyBE2019MNRAS.484.2000D} from the selection of giants included in the {\it GALAH} survey.

In this paper, the abundances of  elements from Li to Eu are used to search for abundance differences between Li-rich and normal giants drawn from the {\it GALAH} survey \citep{BuderGalahDR22018}. Stars are chosen by luminosity and metallicity. Discussion is limited to giants of the Galactic disk, i.e., [Fe/H] $\geq -1$. To minimize intrusion of systematic errors, samples of normal and Li-rich giants are compared for common intervals of luminosity and metallicity. Sample selections and their compositions are discussed in Section \ref{sec:sample} and \ref{sec:analysis}, respectively. This is the first such comprehensive examination of the composition of a large sample of Li-rich giants and an even larger sample of thoroughly otherwise comparable normal giants. In Section \ref{sec:properties}, we discuss various properties (like, mass, rotational velocity, etc.) of Li-enriched giants relative to the Li-normal giants. A discussion of possible origins of Li-rich stars is provided in Section \ref{CreationofLiRichGiants}. Section \ref{sec:conclusion} concludes the paper with some ideas for further advancing understanding of Li-rich giants.

\section{Sampling the H-R Diagram}\label{sec:sample}

\subsection{The {\it GALAH} survey}
The Galactic Archaeology survey with the HERMES spectrograph ({\it GALAH}) is a large-scale spectroscopic survey. The stellar spectra are obtained with the HERMES spectrograph mounted on the 4-m Anglo-Australian Telescope (AAT) at a spectral resolution of 28,000. In the second data release of the {\it GALAH} survey (hereafter, {\it GALAH} DR2), the {\it GALAH} team has provided stellar parameters (Teff, log{\it g}, micro-turbulence velocity) along with quantitatively derived abundances of  23 elements, from Li to Eu for 342,682 stars \citep[see][]{BuderGalahDR22018}. Except for Li, O, Al and Fe, the abundances are based on the assumption of Local Thermodynamic Equilibrium (LTE). Non-LTE Li abundances are obtained from recipes provided by \citep{LindAsplundBarklem2009A&A...503..541L}.
O abundances are estimated from the O triplet (O I 7772 \AA, 7774 \AA, and 7775 \AA) using 1D non-LTE corrections based on the model atom and non-LTE radiative transfer code described in \cite{AmarsiAsplundCollet2015MNRAS.454L..11A,AmarsiAsplundCollet2016MNRAS.455.3735A}, but using 1D {\scriptsize{MARCS}} model atmospheres.
Al abundance are measured from the Al\,{\sc I} lines 6696.023, 6698.673, 7835.309  and 7836.134 \AA \  using 1D non-LTE corrections by \cite{NordlanderLind2017A&A...607A..75N}.
Errors corresponding to each of the given spectral parameters and the elemental abundances are  given. An index Flag$_{cannon}$ against each star indicates the level of confidence in the derived stellar parameters is also provided. Cannon is the name of the reduction software. The Flag$_{cannon}$ index ranges from 0 to 128 with 0 being the most reliable and other numbers indicating issues such as binary companion, fitting accuracies, S/N ratio, etc. (for more details see \cite{BuderGalahDR22018}). 
Similarly, an index $\rm Flag_{abundance}$ corresponding to each element of each star is given. $\rm Flag_{abundance}$ is an indicator of the confidence in measured abundance against each star. The value of $\rm Flag_{abundance}$ index ranges from 0 to 9, with 0 being the most reliable, flag one is raised when line strength is below 2$\sigma$, flag two is raised when the fitting cannon software required an extrapolation, flag three indicates that both flag one and two are raised, flag four is when the $\chi^2$ of best fitting model spectrum is very high or low  and the remaining flags are combinations of these flags \citep[see][for more details]{BuderGalahDR22018}.

For the current study, we start with the red giant sample of \cite{DeepakReddyBE2019MNRAS.484.2000D}, which  includes  51,982 low mass ({\it M} $\leq$ 2 M$_{\odot}$)  stars with reliable atmospheric parameters (i.e., Flag$_{cannon}$ = 0). The  condition Flag$_{cannon}$ = 0 ensures that to a high probability  that none of the considered giant is a binary star \citep{BuderGalahDR22018}. In the HR-diagram, the chosen stars occupy the space defined by 3800K $\leq \rm$ Teff$_{\rm GALAH} \leq$ 5200K and  $0.3 \leq$  $\rm log(L/L_\odot)$ $\leq$ 3.2 dex. This space includes giants from the base of the red giant branch (RGB) through its tip to the  red clump  and the  early asymptotic giant branch (AGB) phase.
All of the selected giants are expected to have reliable atmospheric parameters (i.e., Flag$_{cannon}$ = 0) and uncertainties in Teff of only 100 K and of parallax ($\rm \sigma_{\pi}$/$\rm \pi$ $\leq$ 0.15). The typical error in absolute luminosity  is about 0.05 dex \citep[see][for more details]{DeepakReddyBE2019MNRAS.484.2000D}. In this sample, there are 335 Li-rich RGB stars with Li abundance, A(Li) $\geq$ 1.80 $\pm$ 0.14 dex  of which 20 are super Li-rich with A(Li) $\geq$ 3.20~dex. 
Specific gravity (log$g$) for giants of RGB sample ranges from 0.45 to 4.10 dex with majority of the Li enriched giants falling in a narrow strip defined by log$g$ from 2.0 to 2.9 dex. Typical error in the log$g$ is about 0.19 dex while maximum error is about 0.22 dex. 
The samples of normal and Li enriched giants span the same intervals in effective temperature.

It would be interesting to understand whether some of the lowest Li abundances are indeed measurements. Unfortunately, we do not have access to the actual stellar spectra to examine. Yet, we made an effort to estimate fraction of giants for which abundances may be upper limits. For this, we estimated detection limit for the Li line 6707.7\AA\ using a relation  $\delta$(EW) = $1.6 \times (\omega\ \delta x)^{0.5}$/SNR \citep{Cayrel1988IAUS..132..345C}. By using signal-to-noise ratio (SNR) of individual stars given in the {\it GALAH} catalogue, pixel size ($\delta x$) as 0.047\AA, and FWHM (w) as 0.24\AA\ for the spectral resolution we obtain $\delta$(EW) $\approx$ 2.5 m\AA.\ 
For this study we adopt Li abundance as measurement if the Li line EW is more than or equal to three times that of $\delta$(EW). By estimating EWs using lowest Li abundances in the sample across Teff range  and comparing with the adopted limit we estimate about 2\% of stars might be upper limits or uncertain, and the rest are measurements.
Required atomic data for Li line for conversion from given abundance to EW is taken from \cite{GhezziCunha2009ApJ...698..451G}. To predict the EW corresponding to the given lowest Li abundance across the sample temperature range, we used the radiative transfer code MOOG 2013 version \citep{Sneden1973PhDT.......180S} and LTE stellar atmospheric models from Kurucz \citep{CastelliKurucz2003IAUS..210P.A20C} using atmospheric parameters given in the {\it GALAH} catalogue.

\subsection{Samples by Luminosity}

To a first approximation, cuts in the H-R diagram by  luminosity and effective temperature sample stars over a range in mass  and metallicity but at  similar evolutionary stages. The H-R diagram in Figure \ref{fig:f1} is broken into four regions. The two seemingly most relevant regions for investigations of Li-rich stars  are outlined in the figure. 
The region bounded by $\rm log(L/L_\odot)$ from 1.5 to 1.9 and $\rm Teff_{GALAH}$ from 4650 K to 5200 K includes giants experiencing He-core burning. This region (here, the red clump or RC sample) contains the majority of the normal , the Li-rich and super Li-rich giants.
Below the RC strip in the figure are stars chosen as possible candidates belonging to the bump on the first ascent of the red giant branch (the LB sample). This is defined in the figure by the luminosity limits 1.3 and 1.5 and effective temperature limits 4500 K and 5000 K. Considering range of mass and metallicity of the sample a small overlap between RC and LB boxes can not be ruled out. However majority of the He-core burning giants are in RC box.
The two remaining selections in the H-R diagram  are (i) stars more luminous than the RC and  LB stars, that is, stars with luminosities between 1.9 and 3.0 (the maximum limit for the figure) (here, labelled as RC$^+$ stars but the sample  likely includes stars on the RGB up to its tip) and (ii) subgiant stars with luminosities between 0.3 and 1.3 (i.e., less luminous than the LB sample, here LB$^-$ stars). Our selection from the {\it GALAH} survey provides the following numbers of giants across the four groups:  5951 (RC$^+$), 17909 (RC), 5518 (LB), and 17670 (LB$^-$) of which 31, 229, 23, and 27 are Li-rich, respectively. These numbers cover all metallicities.\\

\begin{figure*}
\includegraphics[width=0.99\textwidth]{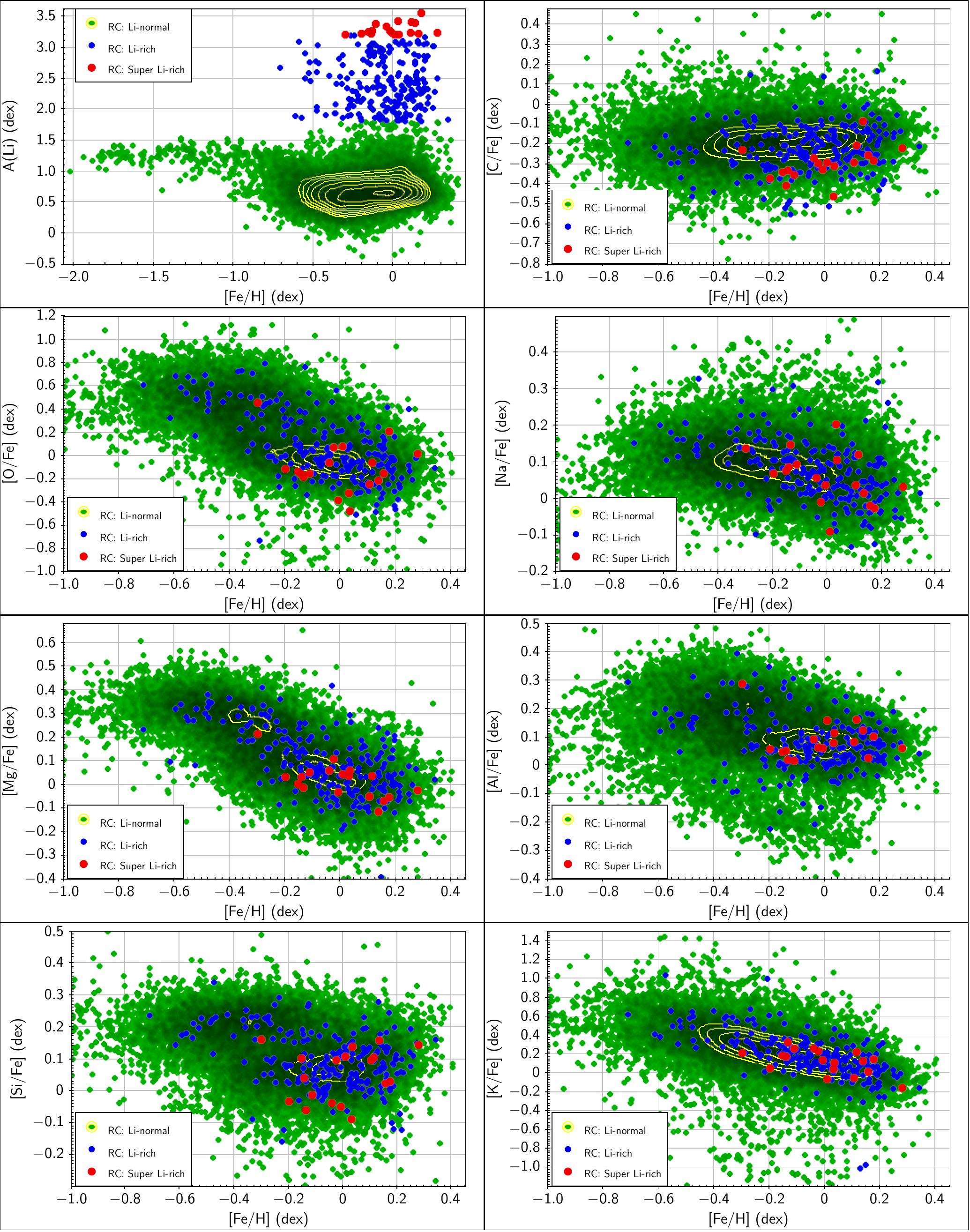}
\caption{Distribution of elemental abundances of Li-normal, rich and super-rich giants of red-clump (RC) sample. The yellow contours in each  panel represent the density of Li normal giants and are provided to ease the visualisation of the position of Li-rich and super rich giants with respect to Li normal giants.
\label{fig:f2a}}
\end{figure*}

\begin{figure*}
\includegraphics[width=1\textwidth]{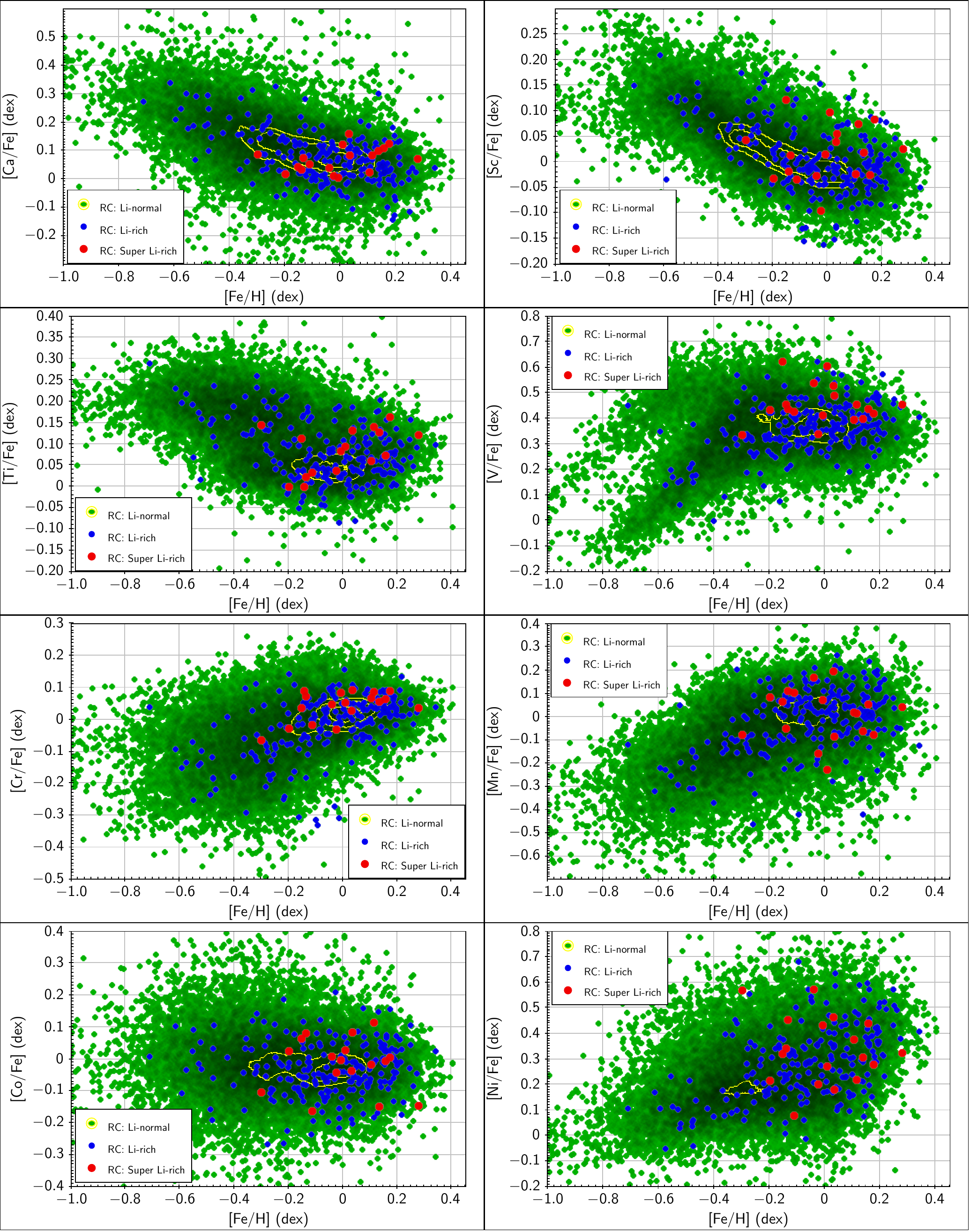}
\contcaption{
\label{fig:f2b}}
\end{figure*}

\begin{figure*}
\includegraphics[width=1\textwidth]{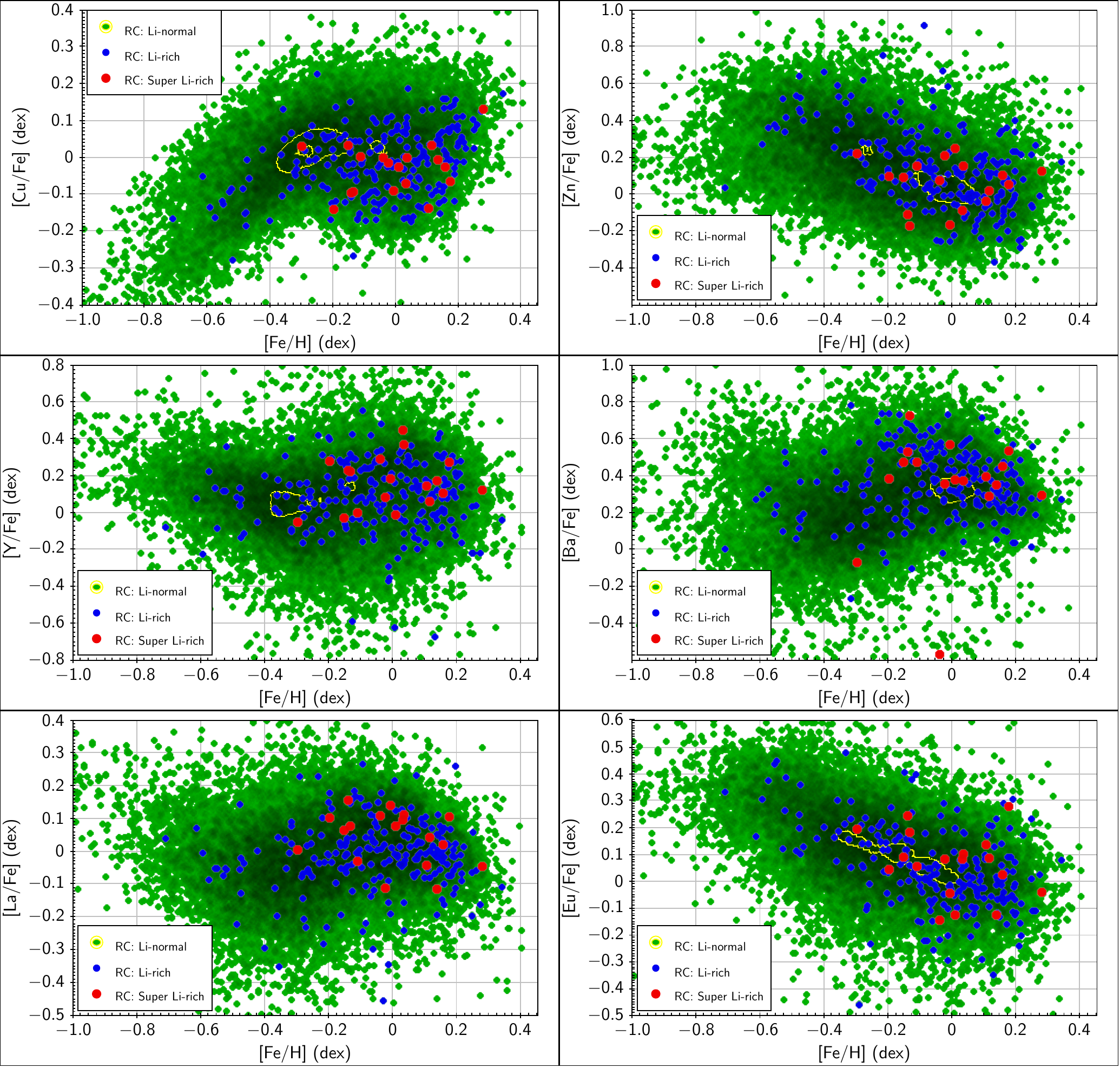}
\contcaption{
\label{fig:f2c}}
\end{figure*}

\begin{table*}
\caption{Mean [Fe/H], A(Li) and [X/Fe] abundances with the standard error of the mean (SEM)  for all the 21 elements given in the {\it GALAH} survey for Li-normal and Li-rich giants of each of the four luminosity samples. Results are tabulated for four [Fe/H] ranges. In the cases where number of stars (data points) is less than five, we have provided the average of the abundance error given in the {\it GALAH} survey, and all such cases are marked with an asterisk (*).
\label{table:t1a}}
\footnotesize
\begin{tabular}{|p{7.2mm}|p{6.5mm}|cccc|cccc|}
    \hline
 & RGB, &   \multicolumn{4}{c|}{ RC$^+$ stars by  [Fe/H] intervals}   &   \multicolumn{4}{c|}{RC stars by [Fe/H] intervals} \\
\cline{3-10}
&Li-& (-0.5, -0.3) & (-0.3, -0.1) & (-0.1, 0.1) & (0.1, 0.3) & (-0.5, -0.3) & (-0.3, -0.1) & (-0.1, 0.1) & (0.1, 0.3) \\
\hline
Stars&normal&1937&1415&538&25&3947&5147&4813&1800\\
&rich&7&13&5&1&18&58&84&59\\
\hline
[Fe/H]&normal&-0.402$\pm$0.001&-0.212$\pm$0.001&-0.031$\pm$0.002&0.133$\pm$0.007&-0.388$\pm$0.001&-0.200$\pm$0.001&-0.004$\pm$0.001&0.168$\pm$0.001\\ 
&rich&-0.441$\pm$0.019&-0.183$\pm$0.014&-0.049$\pm$0.015&0.137$\pm$0.054*&-0.392$\pm$0.015&-0.186$\pm$0.008&0.002$\pm$0.006&0.172$\pm$0.006\\ 
\hline
A(Li)&normal&0.116$\pm$0.010&0.203$\pm$0.012&0.288$\pm$0.020&0.557$\pm$0.116&0.632$\pm$0.003&0.650$\pm$0.003&0.714$\pm$0.003&0.723$\pm$0.005\\
&rich&2.550$\pm$0.175&2.236$\pm$0.090&2.278$\pm$0.169&1.893$\pm$0.117*&2.397$\pm$0.089&2.537$\pm$0.061&2.408$\pm$0.049&2.553$\pm$0.058\\
\hline
[C/Fe]&normal&-0.198$\pm$0.002&-0.186$\pm$0.002&-0.184$\pm$0.004&-0.124$\pm$0.037&-0.203$\pm$0.002&-0.197$\pm$0.001&-0.186$\pm$0.001&-0.167$\pm$0.002\\ 
&rich&-0.286$\pm$0.026&-0.265$\pm$0.021&-0.190$\pm$0.044&-0.183$\pm$0.019*&-0.240$\pm$0.031&-0.262$\pm$0.017&-0.221$\pm$0.012&-0.207$\pm$0.014\\ 
\hline
[O/Fe]&normal&0.367$\pm$0.008&0.164$\pm$0.009&0.021$\pm$0.012&-0.067$\pm$0.056&0.358$\pm$0.004&0.170$\pm$0.004&-0.007$\pm$0.003&-0.073$\pm$0.005\\ 
&rich&0.281$\pm$0.118&0.005$\pm$0.073&-0.385$\pm$0.165&-0.186$\pm$0.082*&0.463$\pm$0.046&0.115$\pm$0.038&-0.018$\pm$0.022&-0.103$\pm$0.025\\ 
\hline
[Na/Fe]&normal&0.108$\pm$0.002&0.094$\pm$0.003&0.088$\pm$0.005&0.138$\pm$0.026&0.113$\pm$0.001&0.095$\pm$0.001&0.071$\pm$0.001&0.052$\pm$0.002\\ 
&rich&0.171$\pm$0.041&0.123$\pm$0.027&0.191$\pm$0.041&0.258$\pm$0.039*&0.148$\pm$0.017&0.111$\pm$0.010&0.060$\pm$0.008&0.033$\pm$0.011\\ 
\hline
[Mg/Fe]&normal&0.190$\pm$0.004&0.065$\pm$0.004&-0.036$\pm$0.007&-0.034$\pm$0.037&0.244$\pm$0.002&0.139$\pm$0.002&0.036$\pm$0.002&-0.009$\pm$0.002\\ 
&rich&0.196$\pm$0.027&0.040$\pm$0.057&-0.167$\pm$0.063&-0.039$\pm$0.061*&0.287$\pm$0.014&0.135$\pm$0.015&0.037$\pm$0.012&0.021$\pm$0.013\\ 
\hline
[Al/Fe]&normal&0.228$\pm$0.003&0.163$\pm$0.003&0.105$\pm$0.004&0.031$\pm$0.021&0.161$\pm$0.002&0.124$\pm$0.002&0.077$\pm$0.001&0.074$\pm$0.002\\ 
&rich&0.277$\pm$0.053&0.159$\pm$0.025&0.055$\pm$0.032&0.081$\pm$0.034*&0.144$\pm$0.034&0.089$\pm$0.016&0.057$\pm$0.009&0.073$\pm$0.007\\ 
\hline
[Si/Fe]&normal&0.203$\pm$0.004&0.150$\pm$0.004&0.132$\pm$0.006&0.074$\pm$0.021&0.156$\pm$0.002&0.115$\pm$0.001&0.078$\pm$0.001&0.094$\pm$0.002\\ 
&rich&0.161$\pm$0.058&0.164$\pm$0.041&-0.014$\pm$0.031&0.099$\pm$0.055*&0.180$\pm$0.023&0.097$\pm$0.013&0.067$\pm$0.008&0.079$\pm$0.010\\ 
\hline
[K/Fe]&normal&0.099$\pm$0.006&0.030$\pm$0.007&-0.025$\pm$0.010&0.042$\pm$0.059&0.359$\pm$0.004&0.232$\pm$0.003&0.113$\pm$0.003&0.005$\pm$0.005\\ 
&rich&0.231$\pm$0.093&0.071$\pm$0.066&0.095$\pm$0.042&0.109$\pm$0.070*&0.466$\pm$0.032&0.277$\pm$0.025&0.140$\pm$0.016&-0.008$\pm$0.030\\ 
\hline
[Ca/Fe]&normal&0.136$\pm$0.002&0.089$\pm$0.002&0.054$\pm$0.004&0.045$\pm$0.024&0.159$\pm$0.002&0.103$\pm$0.001&0.064$\pm$0.001&0.045$\pm$0.002\\ 
&rich&0.166$\pm$0.024&0.114$\pm$0.022&0.080$\pm$0.013&0.058$\pm$0.042*&0.170$\pm$0.020&0.088$\pm$0.011&0.053$\pm$0.007&0.042$\pm$0.010\\ 
\hline
[Sc/Fe]&normal&0.121$\pm$0.002&0.072$\pm$0.002&0.013$\pm$0.003&-0.071$\pm$0.017&0.075$\pm$0.001&0.025$\pm$0.001&-0.014$\pm$0.001&-0.030$\pm$0.001\\ 
&rich&0.181$\pm$0.025&0.058$\pm$0.022&-0.028$\pm$0.015&-0.048$\pm$0.035*&0.076$\pm$0.011&0.032$\pm$0.008&-0.022$\pm$0.007&-0.008$\pm$0.006\\ 
\hline
[Ti/Fe]&normal&0.200$\pm$0.002&0.147$\pm$0.002&0.108$\pm$0.003&0.045$\pm$0.011&0.151$\pm$0.001&0.104$\pm$0.001&0.069$\pm$0.001&0.073$\pm$0.001\\ 
&rich&0.229$\pm$0.026&0.131$\pm$0.023&0.069$\pm$0.014&0.023$\pm$0.028*&0.147$\pm$0.015&0.079$\pm$0.009&0.053$\pm$0.006&0.076$\pm$0.006\\ 
\hline
[V/Fe]&normal&0.433$\pm$0.002&0.404$\pm$0.002&0.359$\pm$0.003&0.296$\pm$0.016&0.326$\pm$0.002&0.381$\pm$0.001&0.371$\pm$0.001&0.351$\pm$0.002\\ 
&rich&0.420$\pm$0.047&0.416$\pm$0.039&0.371$\pm$0.033&0.212$\pm$0.043*&0.284$\pm$0.032&0.349$\pm$0.013&0.393$\pm$0.010&0.392$\pm$0.009\\ 
\hline
[Cr/Fe]&normal&-0.047$\pm$0.002&-0.017$\pm$0.002&-0.001$\pm$0.003&0.004$\pm$0.016&-0.094$\pm$0.002&-0.047$\pm$0.001&0.008$\pm$0.001&0.024$\pm$0.001\\ 
&rich&-0.004$\pm$0.027&-0.034$\pm$0.015&0.036$\pm$0.021&-0.039$\pm$0.042*&-0.098$\pm$0.026&-0.046$\pm$0.013&-0.002$\pm$0.010&0.034$\pm$0.006\\ 
\hline
[Mn/Fe]&normal&-0.206$\pm$0.004&-0.143$\pm$0.005&-0.080$\pm$0.007&0.022$\pm$0.030&-0.136$\pm$0.002&-0.060$\pm$0.002&-0.009$\pm$0.002&-0.002$\pm$0.003\\ 
&rich&-0.213$\pm$0.062&-0.162$\pm$0.032&-0.016$\pm$0.061&-0.001$\pm$0.047*&-0.141$\pm$0.029&-0.025$\pm$0.017&0.018$\pm$0.014&0.018$\pm$0.015\\ 
\hline
[Co/Fe]&normal&0.074$\pm$0.002&0.042$\pm$0.003&0.010$\pm$0.003&-0.084$\pm$0.017&-0.013$\pm$0.002&-0.017$\pm$0.001&-0.026$\pm$0.001&-0.028$\pm$0.002\\ 
&rich&0.070$\pm$0.026&0.004$\pm$0.024&-0.053$\pm$0.014&-0.091$\pm$0.038*&-0.050$\pm$0.039&-0.054$\pm$0.015&-0.042$\pm$0.009&-0.017$\pm$0.010\\ 
\hline
[Ni/Fe]&normal&0.217$\pm$0.003&0.233$\pm$0.003&0.261$\pm$0.006&0.347$\pm$0.029&0.203$\pm$0.002&0.240$\pm$0.002&0.276$\pm$0.002&0.351$\pm$0.003\\ 
&rich&0.174$\pm$0.031&0.222$\pm$0.038&0.223$\pm$0.065&0.217$\pm$0.049*&0.170$\pm$0.034&0.259$\pm$0.016&0.305$\pm$0.015&0.364$\pm$0.016\\ 
\hline
[Cu/Fe]&normal&0.071$\pm$0.002&0.057$\pm$0.003&0.013$\pm$0.004&-0.046$\pm$0.023&-0.017$\pm$0.001&0.029$\pm$0.001&0.020$\pm$0.001&0.054$\pm$0.002\\ 
&rich&-0.034$\pm$0.020&0.048$\pm$0.034&-0.051$\pm$0.047&-0.216$\pm$0.045*&-0.048$\pm$0.020&-0.019$\pm$0.012&-0.028$\pm$0.009&0.023$\pm$0.011\\ 
\hline
[Zn/Fe]&normal&0.068$\pm$0.006&-0.035$\pm$0.007&-0.069$\pm$0.011&0.089$\pm$0.061&0.271$\pm$0.003&0.160$\pm$0.003&0.048$\pm$0.003&0.029$\pm$0.004\\ 
&rich&-0.041$\pm$0.138&0.012$\pm$0.100&-0.149$\pm$0.095&-0.023$\pm$0.058*&0.448$\pm$0.030&0.189$\pm$0.023&0.090$\pm$0.023&0.040$\pm$0.022\\ 
\hline
[Y/Fe]&normal&0.071$\pm$0.005&0.148$\pm$0.006&0.287$\pm$0.010&0.312$\pm$0.050&0.046$\pm$0.003&0.050$\pm$0.003&0.119$\pm$0.003&0.106$\pm$0.005\\ 
&rich&0.062$\pm$0.055&0.150$\pm$0.048&0.230$\pm$0.055&0.278$\pm$0.063*&0.069$\pm$0.033&0.108$\pm$0.023&0.100$\pm$0.022&0.077$\pm$0.024\\ 
\hline
[Ba/Fe]&normal&0.105$\pm$0.006&0.243$\pm$0.007&0.360$\pm$0.010&0.510$\pm$0.071&0.186$\pm$0.004&0.276$\pm$0.003&0.351$\pm$0.003&0.296$\pm$0.004\\ 
&rich&0.235$\pm$0.233&0.203$\pm$0.182&0.577$\pm$0.108&0.592$\pm$0.073*&0.246$\pm$0.223&0.394$\pm$0.230&0.390$\pm$0.189&0.308$\pm$0.018\\ 
\hline
[La/Fe]&normal&0.045$\pm$0.003&0.081$\pm$0.003&0.101$\pm$0.004&0.018$\pm$0.035&-0.054$\pm$0.002&-0.011$\pm$0.002&0.023$\pm$0.002&-0.019$\pm$0.002\\ 
&rich&0.163$\pm$0.046&0.064$\pm$0.025&0.064$\pm$0.040&0.020$\pm$0.038*&-0.058$\pm$0.031&0.007$\pm$0.015&-0.006$\pm$0.015&-0.020$\pm$0.011\\ 
\hline
[Eu/Fe]&normal&0.257$\pm$0.004&0.177$\pm$0.004&0.106$\pm$0.006&-0.039$\pm$0.044&0.173$\pm$0.002&0.103$\pm$0.002&0.036$\pm$0.002&-0.005$\pm$0.003\\ 
&rich&0.468$\pm$0.079&0.146$\pm$0.036&-0.032$\pm$0.044&0.053$\pm$0.043*&0.204$\pm$0.024&0.094$\pm$0.019&-0.021$\pm$0.015&0.009$\pm$0.018\\ 
\hline
\end{tabular}
\end{table*}

\begin{table*}
\contcaption{
\label{table:t1b}}
\footnotesize	
\begin{tabular}{|p{7.2mm}|p{6.5mm}|cccc|cccc|}
    \hline
 & RGB, &   \multicolumn{4}{c|}{LB stars by [Fe/H] intervals}   &   \multicolumn{4}{c|}{LB$^-$ stars by  [Fe/H] intervals}\\
\cline{3-10}
&Li-& (-0.5, -0.3) & (-0.3, -0.1) & (-0.1, 0.1) & (0.1, 0.3) & (-0.5, -0.3) & (-0.3, -0.1) & (-0.1, 0.1) & (0.1, 0.3) \\
\hline
Stars&normal&1229&1590&1592&481&2663&4498&5511&3298\\ 
&rich&0&5&5&7&0&1&5&12\\ 
\hline
[Fe/H]&normal&-0.388$\pm$0.002&-0.197$\pm$0.001&-0.008$\pm$0.001&0.176$\pm$0.002&-0.385$\pm$0.001&-0.192$\pm$0.001&-0.002$\pm$0.001&0.180$\pm$0.001\\ 
&rich&---&-0.192$\pm$0.026&0.029$\pm$0.019&0.212$\pm$0.024&---&-0.202$\pm$0.079*&-0.004$\pm$0.018&0.167$\pm$0.014\\ 
\hline
A(Li)&normal&0.617$\pm$0.006&0.651$\pm$0.005&0.671$\pm$0.005&0.676$\pm$0.010&0.979$\pm$0.005&1.000$\pm$0.004&1.015$\pm$0.003&1.010$\pm$0.004\\
&rich&---&2.390$\pm$0.182&2.392$\pm$0.203&2.408$\pm$0.148&---&1.916$\pm$0.132*&2.012$\pm$0.171&2.186$\pm$0.150\\
\hline
[C/Fe]&normal&-0.177$\pm$0.003&-0.184$\pm$0.003&-0.177$\pm$0.002&-0.162$\pm$0.005&-0.113$\pm$0.003&-0.147$\pm$0.002&-0.158$\pm$0.002&-0.156$\pm$0.002\\ 
&rich&---&-0.252$\pm$0.054&-0.197$\pm$0.021&-0.152$\pm$0.036&---&-0.200$\pm$0.027*&-0.160$\pm$0.048&-0.213$\pm$0.030\\
\hline
[O/Fe]&normal&0.377$\pm$0.007&0.188$\pm$0.007&0.020$\pm$0.006&-0.045$\pm$0.011&0.314$\pm$0.004&0.142$\pm$0.003&0.004$\pm$0.003&-0.065$\pm$0.004\\ 
&rich&---&0.306$\pm$0.111&-0.217$\pm$0.072&-0.261$\pm$0.095&---&0.248$\pm$0.114*&0.070$\pm$0.079&-0.109$\pm$0.036\\ 
\hline
[Na/Fe]&normal&0.108$\pm$0.002&0.067$\pm$0.002&0.036$\pm$0.002&0.029$\pm$0.004&0.089$\pm$0.001&0.042$\pm$0.001&0.003$\pm$0.001&0.019$\pm$0.001\\ 
&rich&---&0.141$\pm$0.038&0.026$\pm$0.017&-0.022$\pm$0.020&---&0.073$\pm$0.053*&0.028$\pm$0.022&0.021$\pm$0.038\\ 
\hline
[Mg/Fe]&normal&0.267$\pm$0.003&0.158$\pm$0.003&0.062$\pm$0.003&0.003$\pm$0.006&0.265$\pm$0.002&0.167$\pm$0.002&0.090$\pm$0.002&0.063$\pm$0.002\\ 
&rich&---&0.152$\pm$0.049&0.041$\pm$0.036&-0.012$\pm$0.031&---&0.355$\pm$0.084*&0.149$\pm$0.028&0.011$\pm$0.024\\ 
\hline
[Al/Fe]&normal&0.235$\pm$0.004&0.179$\pm$0.003&0.114$\pm$0.003&0.090$\pm$0.004&0.192$\pm$0.003&0.127$\pm$0.002&0.082$\pm$0.001&0.083$\pm$0.002\\ 
&rich&---&0.100$\pm$0.021&-0.003$\pm$0.032&0.013$\pm$0.026&---&0.083$\pm$0.047*&0.124$\pm$0.043&0.053$\pm$0.013\\ 
\hline
[Si/Fe]&normal&0.188$\pm$0.003&0.139$\pm$0.003&0.108$\pm$0.003&0.124$\pm$0.004&0.147$\pm$0.002&0.092$\pm$0.002&0.055$\pm$0.001&0.070$\pm$0.002\\ 
&rich&---&0.121$\pm$0.047&0.027$\pm$0.086&0.105$\pm$0.036&---&0.161$\pm$0.077*&0.073$\pm$0.017&0.020$\pm$0.019\\ 
\hline
[K/Fe]&normal&0.220$\pm$0.007&0.116$\pm$0.006&0.014$\pm$0.006&-0.020$\pm$0.012&0.243$\pm$0.004&0.131$\pm$0.003&0.005$\pm$0.003&-0.085$\pm$0.004\\ 
&rich&---&0.246$\pm$0.041&-0.006$\pm$0.022&-0.075$\pm$0.153&---&-0.256$\pm$0.108*&0.027$\pm$0.062&0.063$\pm$0.086\\ 
\hline
[Ca/Fe]&normal&0.181$\pm$0.003&0.117$\pm$0.002&0.070$\pm$0.002&0.032$\pm$0.005&0.180$\pm$0.002&0.109$\pm$0.002&0.055$\pm$0.001&0.038$\pm$0.002\\ 
&rich&---&0.154$\pm$0.019&0.066$\pm$0.027&0.002$\pm$0.030&---&0.244$\pm$0.059*&0.046$\pm$0.033&-0.019$\pm$0.026\\ 
\hline
[Sc/Fe]&normal&0.126$\pm$0.002&0.098$\pm$0.002&0.067$\pm$0.002&0.023$\pm$0.003&0.117$\pm$0.001&0.083$\pm$0.001&0.056$\pm$0.001&0.064$\pm$0.001\\ 
&rich&---&-0.001$\pm$0.016&-0.056$\pm$0.022&-0.047$\pm$0.033&---&0.116$\pm$0.050*&0.087$\pm$0.022&0.020$\pm$0.013\\ 
\hline
[Ti/Fe]&normal&0.197$\pm$0.002&0.148$\pm$0.002&0.118$\pm$0.002&0.103$\pm$0.003&0.182$\pm$0.002&0.128$\pm$0.001&0.092$\pm$0.001&0.090$\pm$0.001\\ 
&rich&---&0.082$\pm$0.055&0.037$\pm$0.034&0.075$\pm$0.037&---&0.164$\pm$0.039*&0.113$\pm$0.070&0.033$\pm$0.024\\
\hline
[V/Fe]&normal&0.482$\pm$0.003&0.469$\pm$0.002&0.438$\pm$0.002&0.384$\pm$0.004&0.443$\pm$0.002&0.455$\pm$0.002&0.450$\pm$0.001&0.439$\pm$0.002\\ 
&rich&---&0.331$\pm$0.041&0.382$\pm$0.037&0.402$\pm$0.019&---&0.515$\pm$0.063*&0.464$\pm$0.062&0.390$\pm$0.034\\ 
\hline
[Cr/Fe]&normal&-0.037$\pm$0.003&0.003$\pm$0.002&0.040$\pm$0.002&0.040$\pm$0.002&-0.020$\pm$0.002&0.008$\pm$0.001&0.039$\pm$0.001&0.059$\pm$0.001\\ 
&rich&---&-0.085$\pm$0.045&0.051$\pm$0.026&0.050$\pm$0.025&---&-0.082$\pm$0.063*&0.001$\pm$0.012&0.025$\pm$0.021\\ 
\hline
[Mn/Fe]&normal&-0.134$\pm$0.004&-0.087$\pm$0.004&-0.042$\pm$0.004&-0.016$\pm$0.008&-0.098$\pm$0.002&-0.039$\pm$0.002&-0.001$\pm$0.002&0.022$\pm$0.002\\ 
&rich&---&-0.001$\pm$0.043&0.130$\pm$0.057&-0.001$\pm$0.060&---&-0.254$\pm$0.066*&0.048$\pm$0.051&0.038$\pm$0.026\\ 
\hline
[Co/Fe]&normal&0.063$\pm$0.003&0.042$\pm$0.003&0.030$\pm$0.003&0.010$\pm$0.004&0.048$\pm$0.003&0.026$\pm$0.002&0.012$\pm$0.002&0.019$\pm$0.002\\ 
&rich&---&-0.078$\pm$0.032&-0.039$\pm$0.030&-0.010$\pm$0.025&---&0.125$\pm$0.054*&0.012$\pm$0.049&0.085$\pm$0.033\\ 
\hline
[Ni/Fe]&normal&0.219$\pm$0.003&0.236$\pm$0.003&0.272$\pm$0.003&0.345$\pm$0.006&0.191$\pm$0.002&0.212$\pm$0.002&0.230$\pm$0.002&0.272$\pm$0.002\\ 
&rich&---&0.167$\pm$0.029&0.372$\pm$0.017&0.359$\pm$0.069&---&0.112$\pm$0.066*&0.223$\pm$0.055&0.250$\pm$0.038\\ 
\hline
[Cu/Fe]&normal&0.081$\pm$0.003&0.110$\pm$0.002&0.099$\pm$0.003&0.115$\pm$0.004&0.056$\pm$0.002&0.089$\pm$0.001&0.091$\pm$0.001&0.128$\pm$0.002\\ 
&rich&---&0.005$\pm$0.029&-0.022$\pm$0.047&0.072$\pm$0.033&---&0.134$\pm$0.063*&0.051$\pm$0.028&0.064$\pm$0.025\\ 
\hline
[Zn/Fe]&normal&0.118$\pm$0.004&0.029$\pm$0.004&-0.035$\pm$0.005&-0.016$\pm$0.01&0.096$\pm$0.003&0.028$\pm$0.002&-0.035$\pm$0.002&-0.041$\pm$0.003\\ 
&rich&---&0.191$\pm$0.097&0.126$\pm$0.104&0.138$\pm$0.064&---&0.153$\pm$0.078*&0.165$\pm$0.067&0.012$\pm$0.044\\ 
\hline
[Y/Fe]&normal&0.017$\pm$0.006&0.032$\pm$0.005&0.108$\pm$0.006&0.126$\pm$0.011&0.005$\pm$0.004&0.030$\pm$0.003&0.074$\pm$0.003&0.055$\pm$0.004\\ 
&rich&---&0.121$\pm$0.066&0.271$\pm$0.090&0.139$\pm$0.103&---&-0.287$\pm$0.088*&-0.068$\pm$0.034&-0.002$\pm$0.068\\ 
\hline
[Ba/Fe]&normal&-0.030$\pm$0.006&0.066$\pm$0.006&0.147$\pm$0.005&0.164$\pm$0.010&-0.053$\pm$0.005&0.026$\pm$0.003&0.034$\pm$0.003&-0.040$\pm$0.003\\ 
&rich&---&0.323$\pm$0.232&0.377$\pm$0.108&0.298$\pm$0.113&---&-0.204$\pm$0.102*&0.057$\pm$0.160&0.139$\pm$0.054\\ 
\hline
[La/Fe]&normal&0.023$\pm$0.004&0.051$\pm$0.003&0.063$\pm$0.003&0.008$\pm$0.005&0.039$\pm$0.003&0.068$\pm$0.002&0.068$\pm$0.002&0.018$\pm$0.003\\ 
&rich&---&-0.118$\pm$0.097&0.030$\pm$0.045&-0.055$\pm$0.053&---&0.065$\pm$0.053*&0.104$\pm$0.045&0.014$\pm$0.031\\ 
\hline
[Eu/Fe]&normal&0.247$\pm$0.004&0.161$\pm$0.004&0.084$\pm$0.004&0.014$\pm$0.007&0.221$\pm$0.003&0.120$\pm$0.003&-0.003$\pm$0.003&-0.107$\pm$0.004\\ 
&rich&---&0.133$\pm$0.059&-0.008$\pm$0.047&0.040$\pm$0.059&---&0.248$\pm$0.058*&0.018$\pm$0.080&-0.139$\pm$0.080\\ 
\hline
\end{tabular}
\end{table*}

\section{Analysis}\label{sec:analysis}

Our search for clues to the origins of Li-rich giants relies on the determination of differences in the abundance ratios [X/Fe] between Li-normal and Li-rich giants in the same evolutionary phase with the same average luminosity, effective temperature and metallicity [Fe/H]. Abundances may also depend on the microturbulence. Within a given luminosity class, the distribution of the microturbulence velocities for the Li-rich, super Li-rich and normal giants are identical. The test is based on the assumption that any systematic errors in abundances will be identical for Li-normal and Li-rich (including the super Li-rich) giants. We divide each of our four evolutionary phases RC$^+$, RC, LB and LB$^-$ into the four metallicity ([Fe/H]) bins $-0.5$ to $-0.3$, $-0.3$ to $-0.1$, $-0.1$ to $0.1$ and $0.1$ to $0.3$ dex centred at $-0.4$, $-0.2$, $0.0$ and $+0.2$ dex, respectively. (Metal-poor ([Fe/H] $< $ $-0.5$) giants are not considered because Li-rich giants at these metallicities are exceedingly rare \citep[see Figure 2 of][]{DeepakReddyBE2019MNRAS.484.2000D}. Their rarity may be due simply to the small samples of metal-poor giants in the {\it GALAH} survey and a low occurrence rate for Li-rich giants, as reported for more metal-rich giants, or to production of Li-rich giants decreasing with decreasing metallicity.) 
For each luminosity sample, relative abundances [X/Fe] at [Fe/H] $= +0.2$, $0.0$, $-0.2$ and $-0.4$ are determined provided  that a sufficient number of abundances are present in the {\it GALAH} database. It is important that Li-rich giants be compared with Li-normal giants of the same luminosity class because different ranges in effective temperature and surface gravity may be involved, for example, inspection of Figure \ref{fig:f1} shows that the RC$^+$ stars are  systematically cooler and of lower surface gravity than the LB stars.
Mean [X/Fe] along with standard error of the mean (SEM) corresponding to each of the 21 elements from C to Eu for both Li-normal and Li-rich populations in each of the luminosity samples are summarized in Table \ref{table:t1a}. 
For Li we have provided the average value of A(Li)
calculated after culling out giants ($\approx$ 2\%) for which provided Li abundances are thought to be upper limits or uncertain,
and for Fe the actual mean value of [Fe/H] corresponding to each  metallicity bin is given. (The  mean values of [Fe/H] offer a check on the difference in the actual mean of two populations in a particular metallicity bin, which can be quite large if distribution along [Fe/H] is not symmetric or homogeneous).

Inspection of Table \ref{table:t1a}  shows that [X/Fe] from O to Eu are the same for Li-normal,  Li-rich and super Li-rich giants within any luminosity class and [Fe/H] interval. The case of C is discussed separately in Section \ref{sec:Li-Carbon}. To illustrate further the lack of a dependence of [X/Fe] on the Li abundance, we show in Figures \ref{fig:f2a} the run of [X/Fe] with [Fe/H] for X = C to Eu for the giants of RC sample. These figures confirm, of course, the numerical results in Table \ref{table:t1a}. The similarity  of [X/Fe] in a given luminosity class and [Fe/H] interval across the Li abundance range serves to challenge theories on how Li enrichment may occur. Two challenges will be mentioned here.

\begin{figure*}
\includegraphics[width=\textwidth]{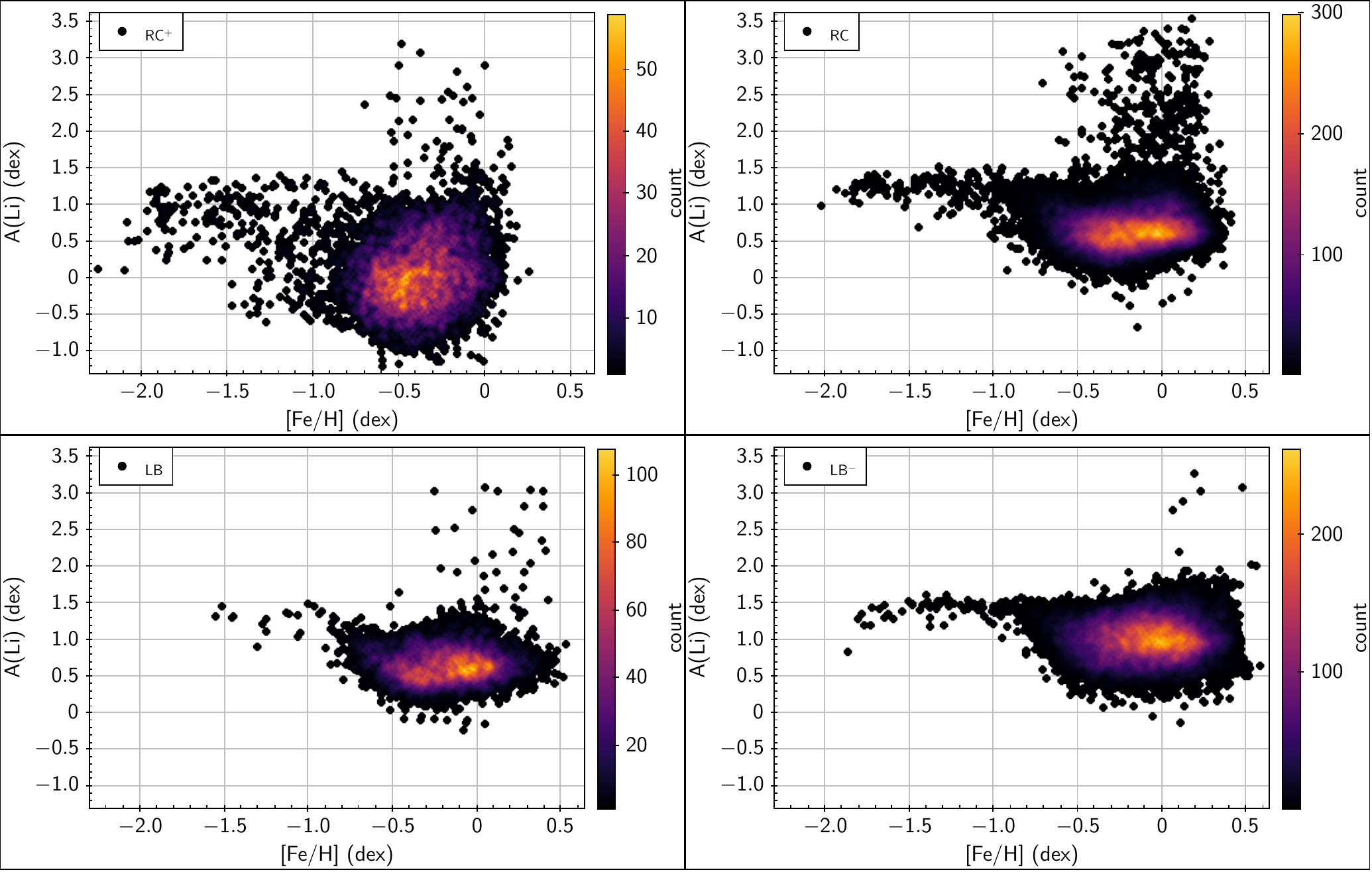}
\caption{Lithium versus [Fe/H] for the {\it GALAH} sample across the four luminosity samples RC$^+$, RC, LB and LB$^-$. \label{fig:f3}}
\end{figure*}

First, there is a  possibility that Li-enrichment might be coupled with  the $s$-process, as in creation of a Li-enriched giant by mass-transfer from a $s$-process and Li-enriched AGB star but this idea is eliminated by the lack of differences in [Y/Fe], [Ba/Fe]  and [La/Fe] between Li-normal and the  Li-rich giants at a given [Fe/H] and for a given luminosity class. Second, accretion of external material (e.g., planets or brown dwarves) may also result in a Li-enriched giant. Accretion may result in abundance changes for some elements in addition to Li enrichment. For example, accretion of Earth-like planets may result in abundance enrichments dependent on the volatility of an element, i.e., on an element's condensation temperature. There is no hint of such a dependence in Table \ref{table:t1a} and Figure \ref{fig:f2a}. For example, in the RC luminosity range and the interval centred on [Fe/H] $= 0.0$, the difference between [X/Fe] for the Li-normal and Li-enriched giants in Table \ref{table:t1a} is within limit $+0.06$ and $-0.07$ for O to Eu with no dependence on an element's condensation temperature. Perhaps, the primary lesson from Table \ref{table:t1a} and Figure \ref{fig:f2a} is that one should look to the simplest theories for an explanation of Li-rich giants (Occam's razor!).

Quite independently of a search for clues to Li enrichment, Table \ref{table:t1a} and  Figure \ref{fig:f2a} are essential tools for mapping the chemical evolution of elements in the thin and thick disk. (Results for the halo are not plotted in this figure.) The Figure \ref{fig:f2a} appears generally consistent with the many surveys now in the literature 
 \citep[e.g.,][]{Reddy2003MNRAS.340..304R,Reddy2006,LiZhaoZhaiJia2018ApJ...860...53L,BuderLindNess2019A&A...624A..19B}.
With respect to the consensus from prior surveys, two kinds of anomalies are apparent in Figure \ref{fig:f2a}. First, the run of [X/Fe] with [Fe/H] for elements shows unexpected structure. Second, the [X/Fe] -- [Fe/H] distribution is offset clearly from previous results. Some elements combine aspects of (i) and (ii).  In category (i), Al and V standout in that there is a run of low [Al/Fe] and [V/Fe] below the main body of the results. In addition but not apparent from the Figure \ref{fig:f2a}, many low values of [Ti/Fe] and [Ba/Fe] have been reported. In category (ii), the obvious cases are again  V as well as  Ni and Ba. These anomalies are discussed fully in the reference paper on the {\it GALAH} observations and their analysis \citep{BuderLindNess2019A&A...624A..19B}. The anomalies of type (i) were excluded in calculating the entries in Table \ref{table:t1a}. Type (ii) offsets are assumed to be unaffected by the Li abundance for a given luminosity and metallicity interval. 
Ti, along with Si and Mg, is one of the most precisely measured element in the {\it GALAH} survey, however due to issues in the wavelength solution of the red channel (covering 6478 -- 6737 \AA) and spikes in the infrared channel (covering 7585 -- 7887 \AA) of the HERMES spectrograph,
several Mg, Si, and Ti abundances show  values below [Mg/Fe], [Si/Fe] and [Ti/Fe] of $-0.3$, and as suggested by \cite{BuderGalahDR22018}, training set distances are the probable reason for these errors and while working with data from the {\it GALAH} survey, one should avoid such affected abundances when possible.
In our study we have excluded all such stars while calculating the mean and standard error of the mean (SEM) for distribution of Mg, Si, Ti and Ba, but have included these stars in evaluating  statistics for other elements with good abundances.

\section{Lithium and other elements}\label{sec:properties}
 
\subsection{Lithium and luminosity}

Investigation of the Li abundances including the distribution of the Li-rich giants with luminosity and metallicity relative to Li-normal giants is possible with the {\it GALAH} sample because whatever selection effects may have entered into the selection of survey stars the effects are unlikely to have biassed the selection  of the Li-rich giants relative to normal giants. 
Distribution of the Li abundances with [Fe/H] for disk giants in the four luminosity samples RC$^+$, RC, LB and LB$^-$ is shown in Figure \ref{fig:f3}. In each luminosity sample, the Li-rich and super Li-rich stars are found at [Fe/H] $> -0.7$ dex. 
In this [Fe/H] range, the occurrence and the frequency of occurrence of Li-rich (including super Li-rich) giants is maximum for the RC sample and apparently increasing with [Fe/H]: the frequency (in \%) is  0.5, 1.1, 1.7 and 3.3 as [Fe/H] increases from the most metal-poor interval centred on $-0.4$, through [Fe/H] $-0.2$ and $0.0$ to the most metal-rich box at [Fe/H] $= +0.2$. The total number of Li-rich giants in the RC sample  with [Fe/H] from $-0.5$ to $+0.3$ dex is 219 out of a Li-normal population of 15707 for a [Fe/H]-independent frequency of 1.4\%. In the most luminous sample RC$^+$, the frequency of occurrence is slightly less than 1\% in each [Fe/H] bin: 26 Li-rich giants in a total sample of 3915 giants for a mean frequency of occurrence of 0.7\%.
Occurrence of Li-rich giants in the LB sample is of considerable interest because of suggestions that internal conditions responsible for the luminosity `bump' may  be ripe for Li production. The total number of Li-normal giants in the LB sample is 4892 with 17 Li-rich stars the frequency of occurrence over the [Fe/H] range in Table \ref{table:t1a} is 0.3\%, a lower frequency than in the RC sample. Finally, the statistics for the subgiant or LB$^-$ sample: 15971  stars and just 18 Li-rich stars for a frequency of occurrence of 0.1\% over the [Fe/H] interval above $-0.5$. There is a clear indication from these frequencies that the phenomenon of Li-rich giants is primarily associated with He-core burning (`clump') giants but as we discuss below these statistics do not necessarily imply that Li production is associated with the He-core burning phase or with the He-core flash immediately preceding  He-core burning in low mass giants.

\begin{figure}
\includegraphics[width=0.5\textwidth]{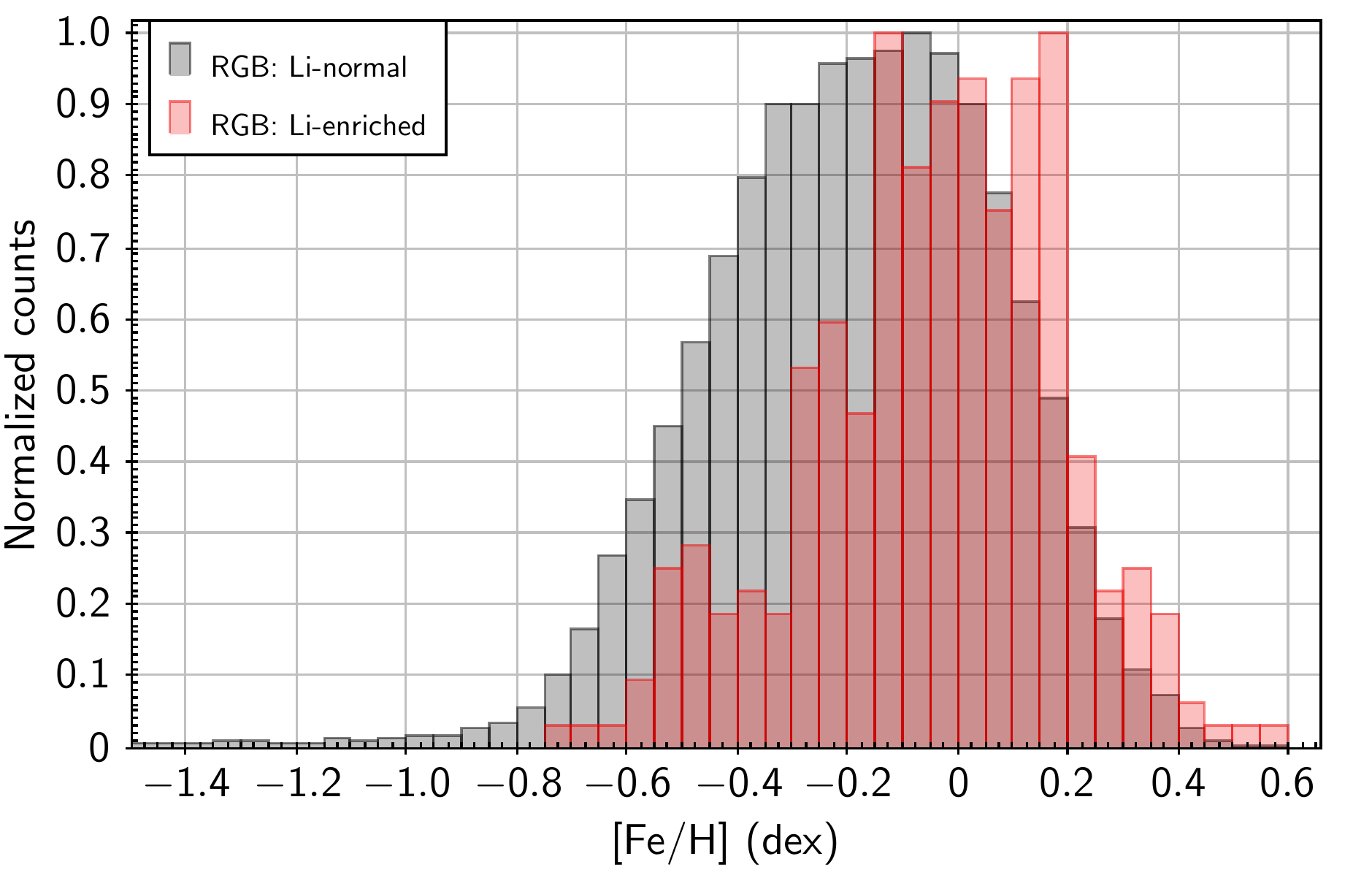}
\caption{Normalized frequency distribution for Li-enriched and normal giants along [Fe/H]-axis.
\label{fig:f4}}
\end{figure}

As an alternative presentation of the [Fe/H]-dependence of the frequency of occurrence for Li-enriched giants, in Figure \ref{fig:f4} we show the frequency distribution of Li-enriched and normal giants along [Fe/H]-axis. Seeing this plot, it is clear that the density distribution for Li-rich giants is shifted by about 0.2 dex towards higher metallicity relative to the distribution for the Li-normal giants, i.e., the frequency of occurrence of Li-rich giants increases with increasing [Fe/H].

Our result that the frequency of occurrence of Li-rich giants increases with increasing [Fe/H] was anticipated by \cite{Casey2019ApJ...880..125C} from the survey of the low-resolution {\it LAMOST} spectra which garnered the staggering total of 2330 Li-rich giants defined as having a lithium abundance A(Li) $\geq 1.5$, a limit lower than our limit by 0.3 dex. Their frequency of occurrence in the interval [Fe/H] $=0.0\pm 0.1$ and over all luminosities was 0.6$\%$. Our estimate for the same [Fe/H] interval and all luminosities is 0.8$\%$ but for A(Li) $\geq 1.8$ in fair agreement with  \cite{Casey2019ApJ...880..125C} despite our tighter definition of a Li-rich giant.

The observed frequency of occurrence of Li-rich giants and its change  with metallicity are quantities which theories about the origins of Li-rich giants should replicate.  A mechanism to produce Li is, of course, the crucial ingredient for any theory but  contributing factors include a star's initial (main sequence) Li abundance, Li depletion (rotational mixing?) on the main sequence and correct modelling of the first dredge-up.  A close quantitative reproduction of the observed frequency of occurrence will be a stiff challenge.

\begin{figure}
\includegraphics[width=0.5\textwidth]{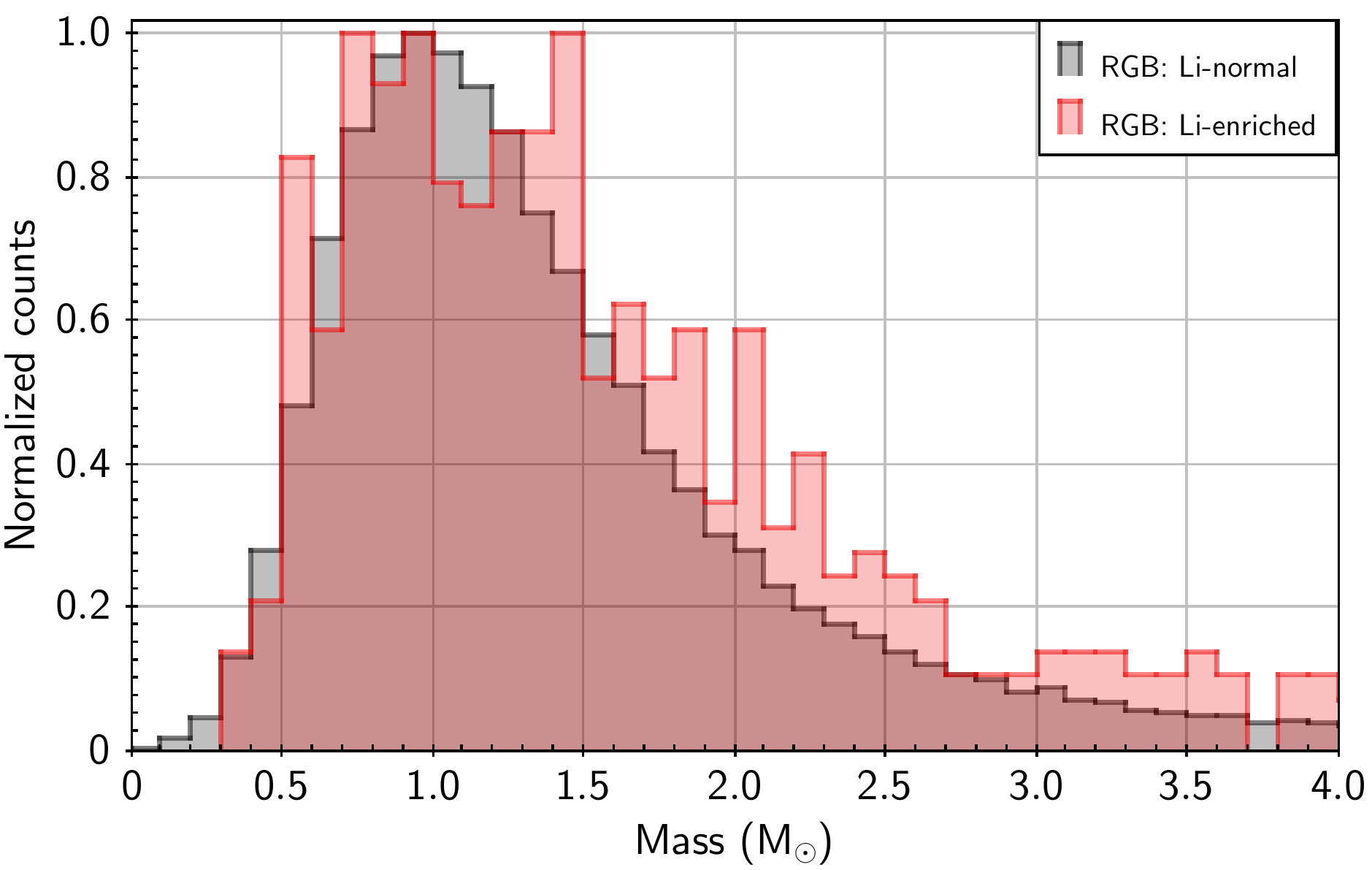}
\caption{Normalized frequency distributions for Li-enriched and normal giants along mass-axis.
\label{fig:f5}}
\end{figure}

\subsection{Lithium and mass}\label{Li-Mass}

There is interest in how the Li-rich giants are distributed in stellar mass. Stellar masses have been estimated, as in \cite{DeepakReddyBE2019MNRAS.484.2000D}, using the formula
{$log(M {\rm /M_\odot}) = {log(L/{\rm L_\odot}) +
log{g} - log{\rm g_\odot} + 4 \times log(\rm Teff_\odot/Teff_{GALAH})}$},
where we used
log${\it g_\odot}$ = 4.44~dex and Teff$_\odot$ = 5772 K for the solar values. Values of effective temperature, Teff$_{\rm GALAH}$, and logarithmic surface gravity, log{\it g}, are adopted from the {\it GALAH} catalog \citep{BuderGalahDR22018} and the values of luminosities are based on parallaxes and apparent magnitudes taken from the Gaia catalogue.
We used bolometric correction based on $\rm Teff_{\rm GALAH}$ and the relation given in \cite{AndraeFouesneauGaiaDR2StellarParametersFromAPSIS2018}.
Values of luminosities given in the Gaia and those derived in this study using $\rm Teff_{\rm GALAH}$ agree well with each other. The difference between the two values corresponds to a  $\sigma\approx$ 0.02~dex.
Uncertainties in the derived luminosity values due to errors in parallax and $\rm Teff_{\rm GALAH}$ are   $\sigma_{L/L_\odot}$ $\leq$ 0.05~dex for most of the stars.

\begin{figure*}
\includegraphics[width=1\textwidth]{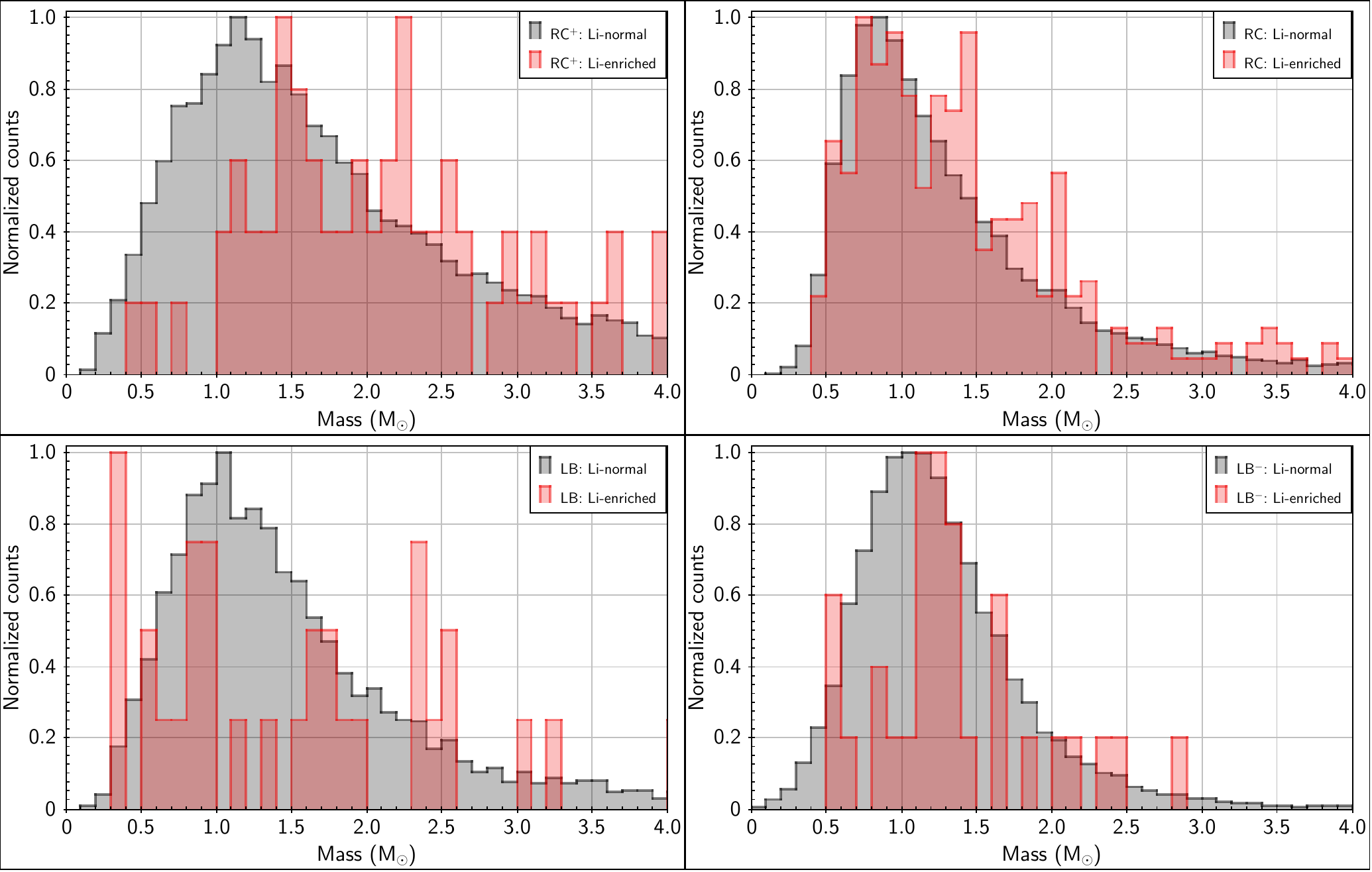}
\caption{Normalized frequency distributions for Li-enriched and normal giants of RC$^+$, RC, LB and LB$^-$ samples along mass-axis.
\label{fig:f6}}
\end{figure*}

Masses of all the 64,493 giants of the {\it GALAH} survey populating the red giant region defined by 3800 $\leq$ $\rm Teff_{GALAH}$ $\leq$ 5200 K and  $0.3 \leq$  $\rm log(L/L_\odot)$ $\leq$ 3.2 dex, and having reliable stellar parameters (Flag$_{cannon} = 0$) along with small uncertainties in effective temperature ($\rm \sigma_{Teff_{GALAH}}\leq$ 100 K) and parallax ($\rm \sigma_{\pi}$/$\rm \pi$ $\leq$ 0.15) are estimated. The typical error in the estimated masses is about 50\%. 
Also, comparison of the stars' positions with theoretical isochrones in the Luminosity--Effective temperature and Surface gravity--Effective temperature  planes, we found that most of the giants in our sample are low mass giants (with $M \leq 2$ M$_\odot$).
In Figure \ref{fig:f5}, we show the normalized frequency distributions for masses of Li-rich and normal giants. Inspection of the figure suggests that the probability of becoming a Li-rich giant is essentially independent of a star's mass. Also, probability of conversion to a Li-rich giant decreases sharply above about 3M$_\odot$ and vanishes above about 6M$_\odot$.

It is of interest to also know if mass distribution of Li-rich and normal giants of RC$^+$, RC, LB and LB$^-$ samples show any difference when considered individually. In Figure \ref{fig:f6}, we have shown the normalized frequency distribution for masses of Li-rich and normal giants for all of the four groups. For all the four groups, spreads in distribution for Li-rich and normal giants along mass axis are similar.
Also, based on the Kolmogorov-Smirnov test (K-S test) we found that the distribution for Li-rich and normal giants of LB$^-$ and LB groups are statistically similar with a significance level of better than 5\%. However, in case of RC$^+$ and RC samples, the K-S test failed to find similarity in the distributions of Li-rich and normal giants with a significance level of 5\% and further favours an alternative hypothesis that the cumulative distribution function (CDF) of Li normal giants is larger than the CDF of Li-rich giants.

Estimated masses from the above equation with their typical uncertainty are in some cases very low ($< 0.6 M_\odot$) which may appear to be unphysical in that the giant's age exceeds that of the {\it Galaxy}. Severe mass loss as a giant may reconcile low mass and age. The referee directed our attention to the giant star KIC 4937011 or W007017 in the open cluster NGC 6819. This giant is Li-rich with A(Li) = 2.3 \citep{Anthony-Twarog2013ApJ...767L..19A} with a location in the H-R diagram below the red boundary of the cluster's red clump. Asteroseismology gives the star a mass  $M = 0.71\pm0.08M_\odot$  and shows that it has the structure of a RC star \citep{HandbergBrogaard2017MNRAS.472..979H}. One may speculate that this star experienced a severe loss of mass at the He-core flash. Mass loss is {\it not} a common event in this cluster: \cite{HandbergBrogaard2017MNRAS.472..979H} from {\it Kepler} asteroseismology  show that the cluster's RGB and RC giants have very similar masses -- $M = 1.61\pm0.02$ $\rm M_\odot$ for RGB giants and $M = 1.64 \pm 0.02$ $\rm M_\odot$ for the RC giants. Thus, KIC 4937011 is an exceptional Li-rich star. Does it have relatives among the {\it GALAH} survey's Li-rich giants?

\begin{figure}
\includegraphics[width=0.5\textwidth]{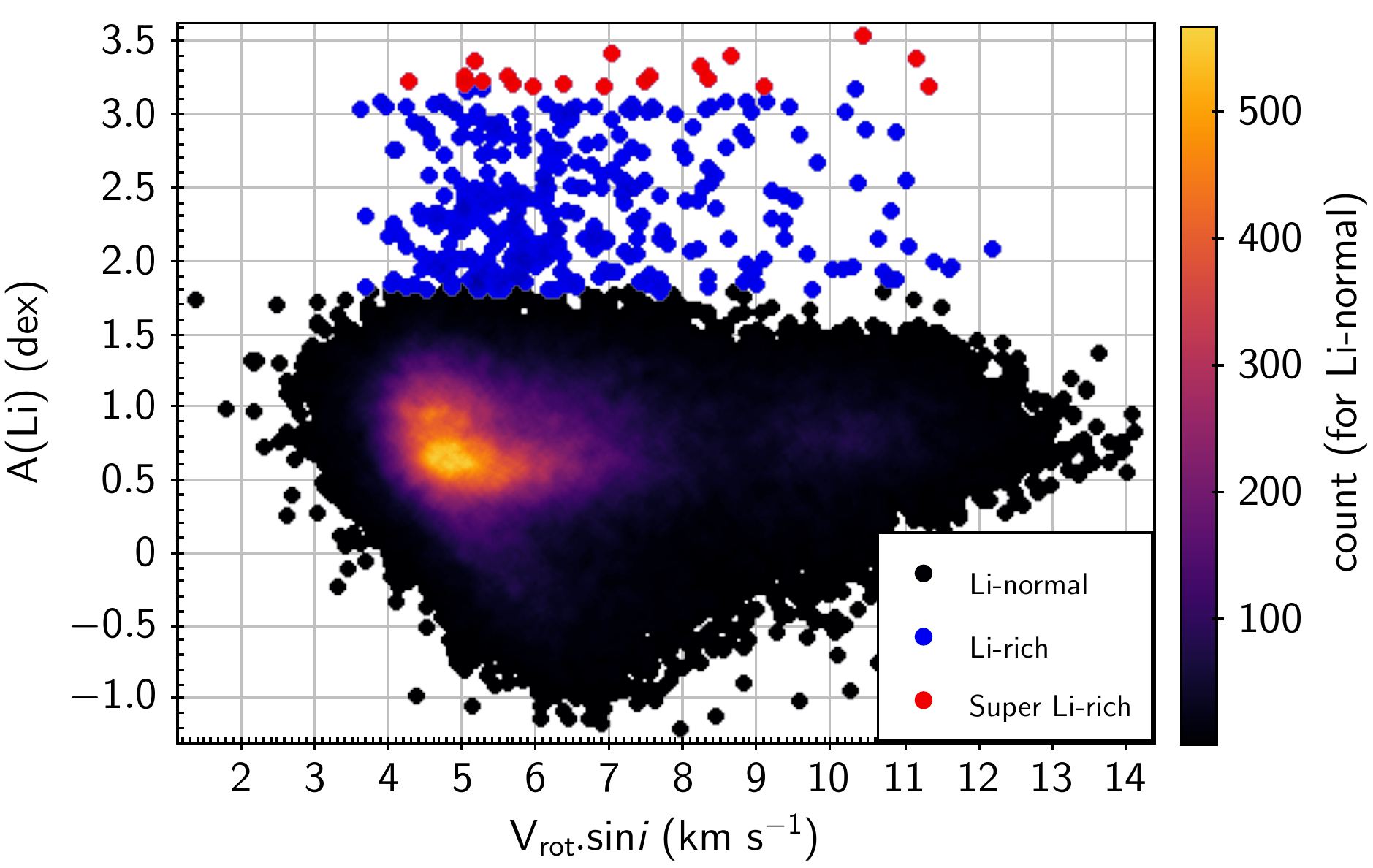}
\caption{A(Li) versus the projected stellar rotational velocity  ($V_{\rm rot} {\rm sin}i$) for our sample of giants from the {\it GALAH} survey.
\label{fig:f7}}
\end{figure}

\begin{figure}
\includegraphics[width=0.48\textwidth]{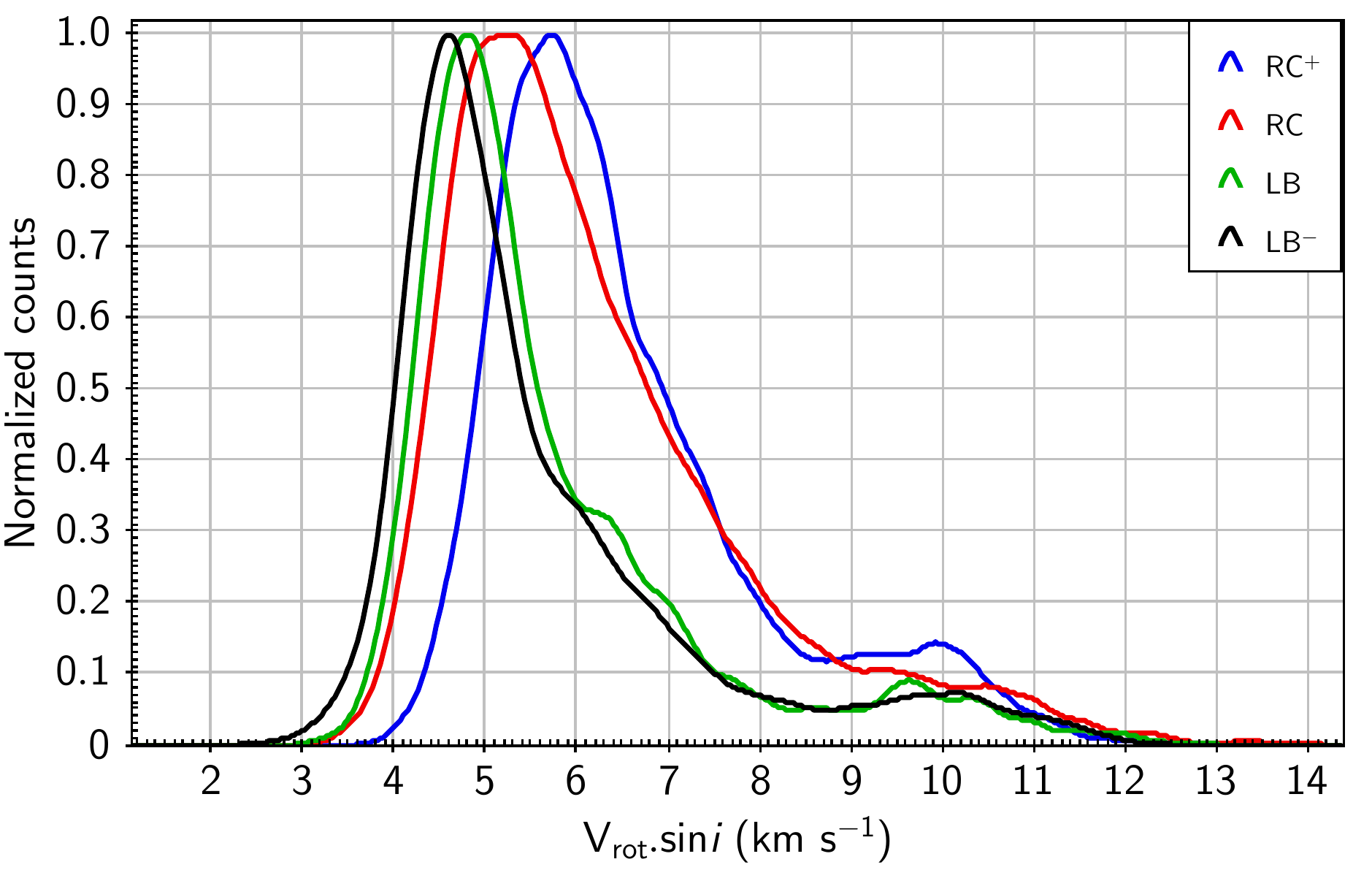}
\caption{Normalized density distribution in projected stellar rotational velocity ($V_{\rm rot} {\rm sin}i$) for our sample of  RC$^+$, RC, LB and LB$^-$ normal giants.
\label{fig:f8}}
\end{figure}

\begin{figure*}
\includegraphics[width=\textwidth]{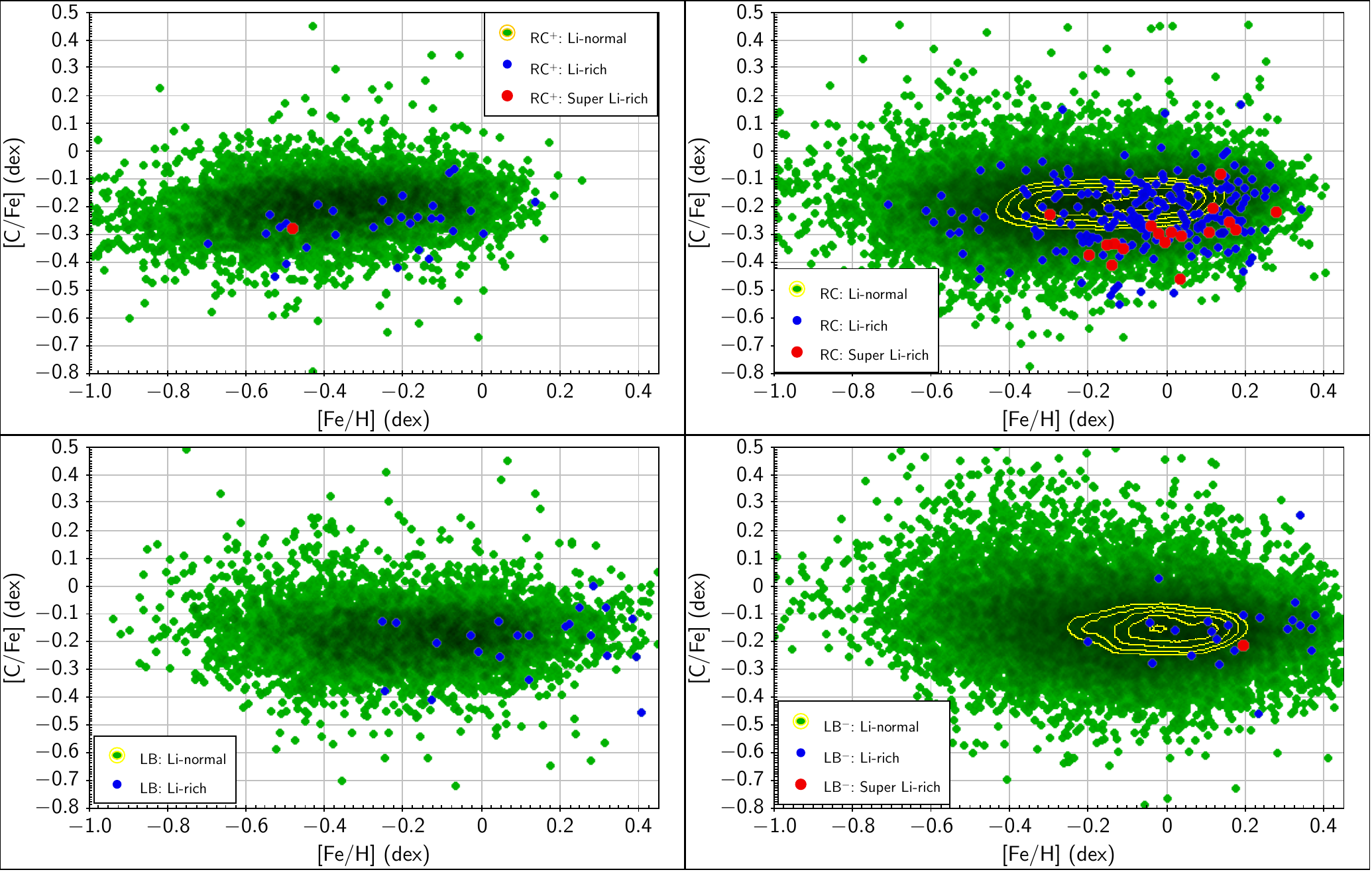}
\caption{[C/Fe]] versus [Fe/H] for Li-normal, rich and super-rich giants of RC$^+$, RC, LB and LB$^-$ samples.
\label{fig:f9}}
\end{figure*}

\begin{figure*}
\includegraphics[width=\textwidth]{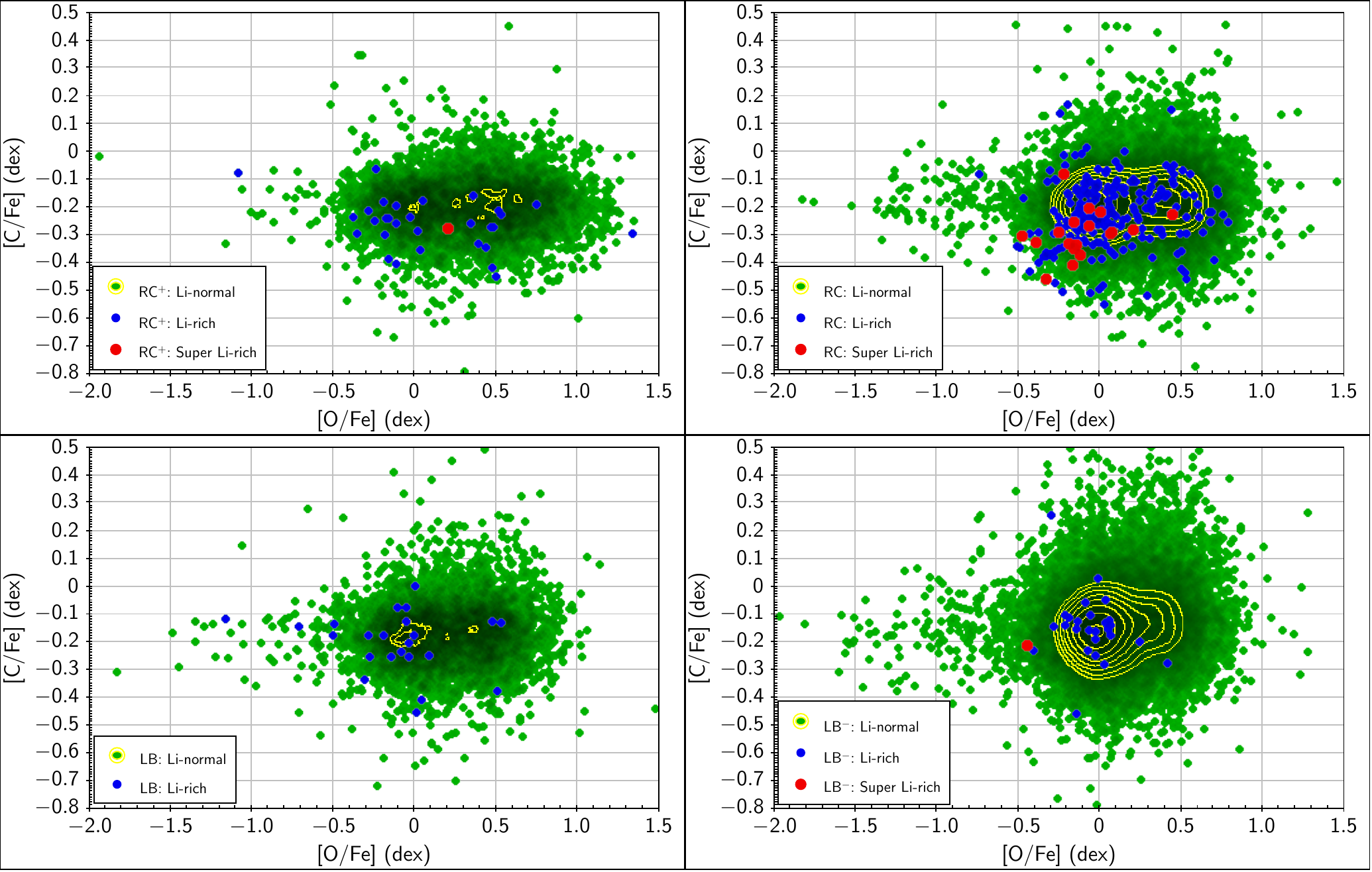}
\caption{[C/Fe] versus [O/Fe] for Li-normal, rich and super-rich giants of RC$^+$, RC, LB and LB$^-$ samples.
\label{fig:f10}}
\end{figure*}

\subsection{Lithium and rotational velocity} \label{sec:Li-Vsini}

It is of interest to know how the Li enriched and normal giants are distributed in stellar rotational velocity as rotational velocities of stars have been linked to Li enhancement scenarios such as mergers \citep{SiessLivio1999MNRAS.308.1133S}, tidal interactions between binary stars  \citep{Casey2019ApJ...880..125C}. 

The {\it GALAH} survey provides the projected rotational velocity  ($V_{\rm p} = V_{\rm rot} sini$, where "$i$" is the inclination angle).
Derived values of $V_{\rm p}$ are based on spectral linewidths which also consists of effects of macro- and micro-turbulence velocities, their combined effect on spectral width is not insignificant. With that caution, we plot the distribution of projected rotational velocity  among red giants of normal and enhanced Li. Average value of $V_{\rm p}$ seems to increase as stars evolve along the RGB as shown in Figure \ref{fig:f8}. Values for RC$^+$ giants, most evolved among our giants, have an average  2 km s$^{-1}$ higher values compared to the least evolved giants of LB$^{-}$.
Given the moderate spectral resolution ($\lambda/\Delta \lambda = 28,000$ or 10 km s$^{-1}$) and the presence of atmospheric (microturbulence and macroturbulence) motions of several km s$^{-1}$, it is sufficient to note that no particular trend of V$_{\rm rot}$ of Li-rich giants compared to Li normal giants (see Figure \ref{fig:f7}) has been noticed.
Although there is no discernible difference between the rotational velocity distributions for normal and Li-rich giants within a given luminosity sample, it is seen from Figure \ref{fig:f1} that with but a single exception the super Li-rich stars belong to the RC sample.

\subsection{Lithium and the Carbon Abundance} \label{sec:Li-Carbon}

Comparison of elemental abundances C to Eu in Li-normal and Li-rich giants is a potentially powerful clue to the origins of the Li enrichment. In Figure \ref{fig:f2a}, abundances for the RC sample are shown. The Li versus [Fe/H] distribution is shown in the top-left panel with the density of points representing Li-normal giants shown by the yellow contours. 
 For these density contours suitable smoothing and scaling (linear or logarithmic) are used depending on the distribution in the individual plots. The purpose of using contours in Figure \ref{fig:f2a} is to  identify the maximum density region for Li normal giants and to  see if occurrence of Li-rich and super Li-rich giants favour any particular [Fe/H] and [X/Fe] (and so  the density value corresponding to each  contour do not need to be specified).
Li-rich and super Li-rich giants are denoted (here and in all panels) by blue and red filled points, respectively.  Other panels of this figure show [X/Fe] versus [Fe/H] where X runs successively from C to Eu. Abundance differences between the normal and Li-rich giants will be seen as a displacement of the Li-rich (blue dots) and/or the super Li-rich (red dots) giants with respect to the normal giants (green dots and the yellow contours) with displacements in one or both of the axes [X/Fe] and [Fe/H]. In an earlier section, we commented on two applications of the [X/Fe] values to test ideas on forming Li-rich giants: a test involving abundances as a function of condensation temperature and another looking for correlation with $s$-process abundances. Here, we consider the carbon abundances.

Inspection of the panel showing [C/Fe] versus [Fe/H] (Figure \ref{fig:f9}) suggests that Li-rich giants and normal giants have the same [C/Fe] deficiency, a deficiency attributed to the first dredge-up. The super Li-rich giants, as noted earlier, have, however, a  systematically lower [C/Fe] than the normal and the Li-rich giants by almost 0.2 dex and this difference in [C/Fe] may increase at lower [Fe/H]. Our sample of super Li-rich giants appears confined to the more metal-rich giants (i.e., the thin disk), say [Fe/H] $\geq -0.2$ with a lone example  at [Fe/H] = $-0.3$. In contrast, the Li-rich giants are spread roughly uniformly over the [Fe/H] range covered by the normal giants -- see, Figure \ref{fig:f9} which shows the [C/Fe] versus [Fe/H] plots for the RC$^+$, RC, LB and LB$^-$ samples. Although the populations of giants in the luminosity samples RC$^+$, LB and LB$^-$ are smaller than the well populated RC sample,  the RC$^+$ sample exhibits the same lower [C/Fe] values for Li enhanced giants as the RC sample:  each of the RC$^+$, LB and LB$^-$ samples show no (or a very slight) difference in [C/Fe] between the normal Li and Li-rich stars.
More obviously and as noted previously, there is an almost complete absence of super Li-rich giants in the RC$^+$, LB and LB$^-$ samples: just one super Li-rich giants in the RC$^+$ and another in the LB sample and one cannot exclude the possibility that these are merely misclassifications which really belong to the RC sample.

The implication of the lower [C/Fe] for the super Li-rich giants may be that prior to or concurrently with becoming a Li rich the giant experiences an episode of internal mixing and additional exposure to the CN-cycle following evolution through the LB$^-$ and LB phases but prior to the RC phase. This episode of mixing brings additional CN-cycled and Li-enriched material to the surface to create the super Li-rich giant. The observation that the Li-rich and normal giants are similarly distributed in the [C/Fe] -- [Fe/H] plane suggests the episode of Li production may be unrelated to the first dredge-up held responsible for the decrease in the C abundance. Destruction of C and its conversion to N by the CN-cycle preserves the sum of the C and N abundances. The sum provides a valuable test of CN-cycling but unfortunately, the {\it GALAH} survey does not provide the N abundance. Note that the {\it GALAH} C abundance is derived from a C\,{\sc i} line with a negligible isotopic wavelength shift and, hence, the C abundance is the sum of the $^{12}$C and $^{13}$C abundances. The $^{12}$C/$^{13}$C ratio provides additional insight into CN-cycling.

If the additional reduction of C in super Li-rich  giants is a consequence of mixing in giants, there is a  possibility that O might be reduced with a concomitant  further increase in the N abundance and  that Na might be enhanced also by a small amount. Nitrogen will be enhanced with the expectation that the sum of the C and N abundances will be conserved. Figure \ref{fig:f2a} for RC giants includes panels for [O/Fe] and [Na/Fe]. Inspection of the panels shows that [O/Fe] for the super Li-rich giants in the thin disk have the [O/Fe] of the normal giants. The same conclusion about [O/Fe] applies to the Li-rich giants in both the thin and thick disks.
However, Figure \ref{fig:f10} suggest a positive correlation between [C/Fe] and [O/Fe] among super Li-rich giants i.e., super Li-rich giants with lower [C/Fe] also have lower [O/Fe]. In this figure, the stars with [O/Fe] $<$ 0.2 dex are thin disk stars while the stars with [O/Fe] $>$ 0.2 dex are thick disk stars. Although the super Li-rich giants are chemically distinct than the normal giants they are well distributed in the kinematical plane \citep{DeepakReddyBE2019MNRAS.484.2000D}.
Similarly, the [Na/Fe] values of super Li-rich and Li-rich giants do not distinguish these giants from the large population of normal giants.  These are not surprising results for O and Na because deep mixing is required to alter their surface abundances and, in addition, the deep mixing may destroy Li and result in a depleted not enhanced  surface Li abundance.

To set the picture expected of C and N abundances, we estimate the effect of conversion of C to N by the CN-cycle using the solar abundances for the initial composition  of a [Fe/H] = 0 giant: A(C)$_\odot$ = 8.43 $\pm$ 0.05,  A(N)$_\odot$  = 7.83 $\pm$ 0.05 and A(O)$_\odot$  = 8.69 $\pm$ 0.05 are solar C, N, and O abundances, respectively, taken from \cite{Asplund2009} which is the same reference as used \cite{BuderGalahDR22018}. Li-rich and normal giants show a C under abundance [C/Fe] $\simeq -0.2$ which if the sum of the solar C and N abundances is conserved translates to [N/Fe] $\simeq +0.4$. Similarly, the super Li-rich giants at [Fe/H] = 0 show [C/Fe] $\simeq -0.3$ and then [N/Fe] $\simeq +0.5$ is expected. Abundance uncertainties in the solar or reference abundances contribute about $\pm0.05$ to these estimates for the N abundance. These estimates assume that the C abundance includes both isotopes $^{12}$C and $^{13}$C. If the C abundance comes from CH lines and is thus primarily $^{12}$C the predicted N abundance might be reduced by about 0.12 dex for $^{12}$C/$^{13}$C $\sim$ 3. If the ON-cycle were to also be involved, the sum of the C, N and O abundances is conserved but, since the solar O abundance is greater than the solar C abundance and far greater than the solar N abundance, larger increases of [N/Fe] occur for even mild conversions of O to N.

Nitrogen as N\,{\sc i} lines are not detectable in spectra of red giants but is readily sensed  through lines of the CN radical via its red and violet systems. (The NH molecule is also detectable at 3360 \AA\ and possibly too via the NH infrared vibration-rotation lines near 3 microns.)  Carbon is also available through CH lines near 4300 \AA. Combining CN and CH lines provides the N abundance but with the potential for an anti correlation between the C and N abundances resulting from systematic errors in the C abundances from CH being transferred to the CN analysis. Also, there is the prospect of elemental abundances from a molecule differing systematically from abundances from atomic lines, especially from high-excitation lines such as the C\,{\sc i} lines; the atomic and molecular lines are formed in different regions (deep versus shallow) of the stellar photosphere. In an attempt to explore the C-N connection for Li-enriched giants, we have considered two ways of determining the N abundance for Li-enriched giants including some in the {\it GALAH} survey: (i) Li-rich giants have been analysed previously by many different investigators for C and O (and many other elements) using high-resolution optical spectra, and (ii) the low-resolution {\it LAMOST} survey of millions of giants covers the blue CH and CN bands and \cite{XiangTingRix2019arXiv190809727X} have calibrated {\it LAMOST} abundances including C and N using abundances from  high-resolution optical ({\it GALAH}) and infrared ({\it APOGEE}) surveys.

\begin{figure*}
\includegraphics[width=1\textwidth]{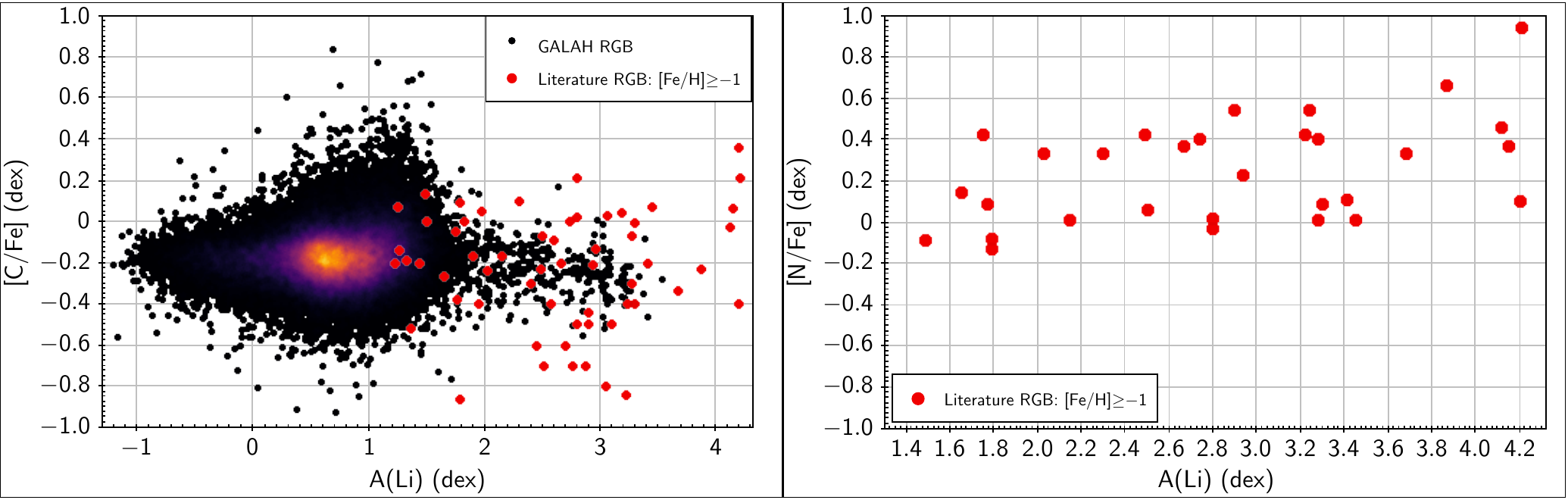}
\caption{Abundances of C and N for known Li-enriched (including a few normal) giants in literature. For carbon, the RGB sample extracted from the {\it GALAH} survey is also plotted in the background.
\label{fig:f11}}
\end{figure*}

In seeking the N abundances of Li-rich giants, we searched the literature for analyses of high-resolution optical spectra and collected Li abundances along with abundances of other elements, as available, as well as stellar parameters.
Data for about 400 giants were compiled. Results for giants which were not Li-rich were included as reported in the papers describing the analysis of Li-rich giants but no special effort was made to include normal giants analyzed similarly by the authors reporting on the Li-rich giants. Selection of giants was restricted to member of the disk, i.e., [Fe/H] $> -1.0$, the same limit as applied to selection of Li-rich giants from the {\it GALAH} survey. 
It is also important to note that many of the Li-enriched giants studied in the literature are not field stars and belong to globular clusters. Depending on the metallicity, etc. of the clusters the MS-turnoff and position of RGB tracks can vary widely along Teff and luminosity axis which can further induce scatter in Li and CNO abundances. To avoid any contamination from such giants, we excluded stars with Teff $>$ 5200 K and log$g$ $>$ 4.1 dex (which is the same limit as for our RGB sample from the {\it GALAH} survey). This limit also make sure that we only have giants from the literature. This resulted in a sample of about 250 giants some of which have been observed and analysed multiple times. 
It is also important to note that studies in the literature have used different atomic and molecular lines to  estimates the C abundances. Some of these lines are highly divergent, for example the most divergent C\,{\sc i} line at 5380 \AA \ yield more than an order of magnitude higher C abundance than that from the Swan system C$_2$ line (5135 \AA) at 4700 K \citep[][and references therein]{LuckHeiter2007}. So in this study, to avoid any contamination, we have excluded the C abundances which are estimated only from the C\,{\sc i} line at 5380 \AA.

\begin{figure}
\includegraphics[width=0.5\textwidth]{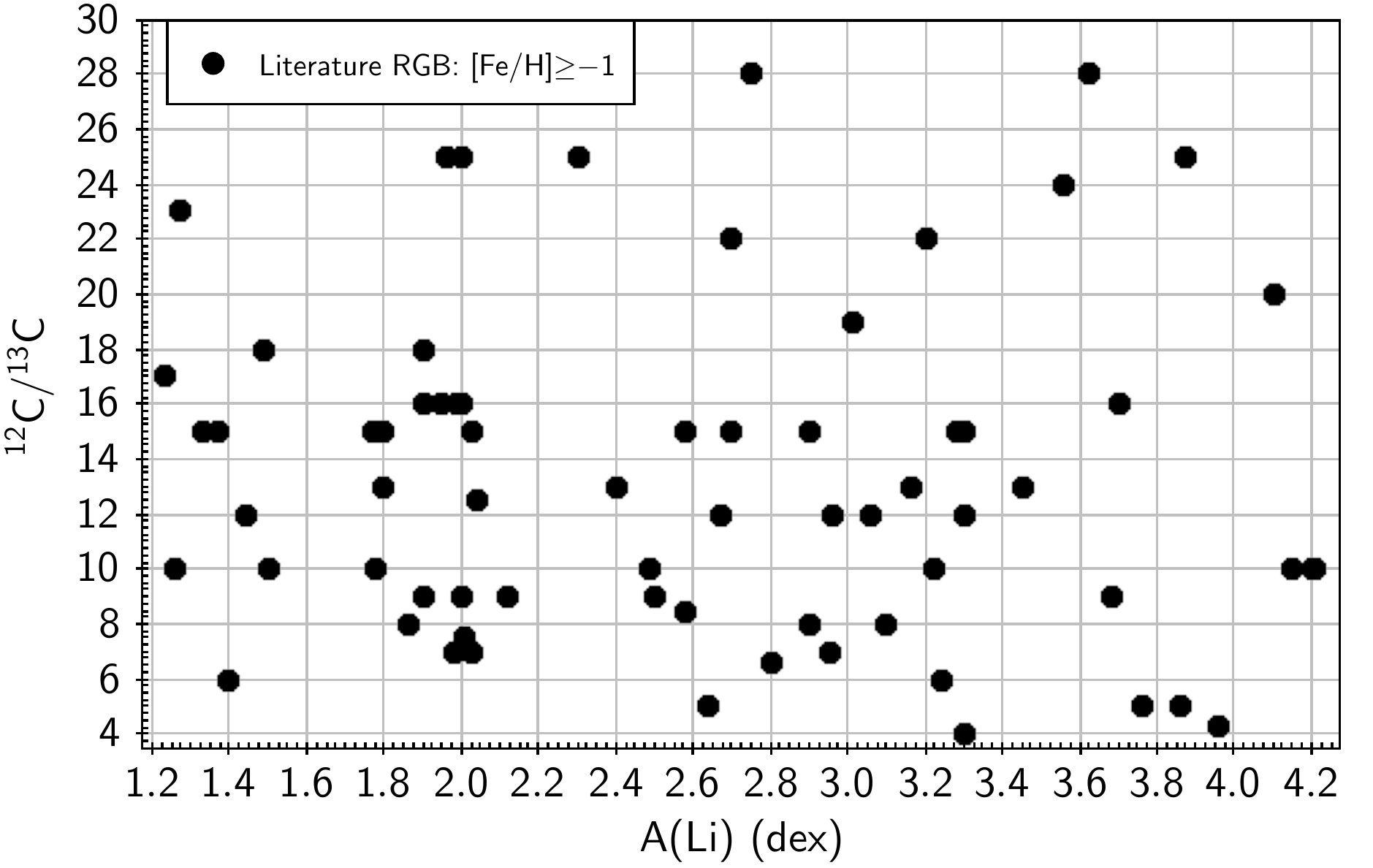}
\caption{$^{12}$C/$^{13}$C versus Li abundances for known Li-enriched (along with a few normal) giants in the literature.
\label{fig:f12}}
\end{figure}

Results from the literature are generally in good accord with those from the {\it GALAH} survey as in Figure \ref{fig:f2a}. A partial presentation is provided in Figure \ref{fig:f11} where [X/Fe] versus A(Li) is provided for X = C and N. Scatter in these plots must partially arise from combining results by different investigators. The few remarkable outliers may also arise from errors in an analysis. Results in Figure \ref{fig:f11} are generally consistent with ours. The novel contribution of Figure \ref{fig:f11} is, of course, the panel providing the results for [N/Fe]. Our expectation  is that [N/Fe] should be close to $+0.4$  if, as seen in Figure \ref{fig:f11}, [C/Fe] is about $-0.2$. In Figure \ref{fig:f11}, [N/Fe] is close to expectation but slightly greater than expected. Note, however, that a systematic overabundance of the C abundance necessarily results in an under abundance of the N abundance thanks to the use of the CH and CN molecules in the C and N abundance analysis. The $^{12}$C/$^{13}$C ratios also taken from the literature are shown in Figure \ref{fig:f12} as a function of A(Li). There is a large spread in the isotopic ratio at all Li abundances from normal  to the super Li-rich giants.

\begin{figure}
\includegraphics[width=0.5\textwidth]{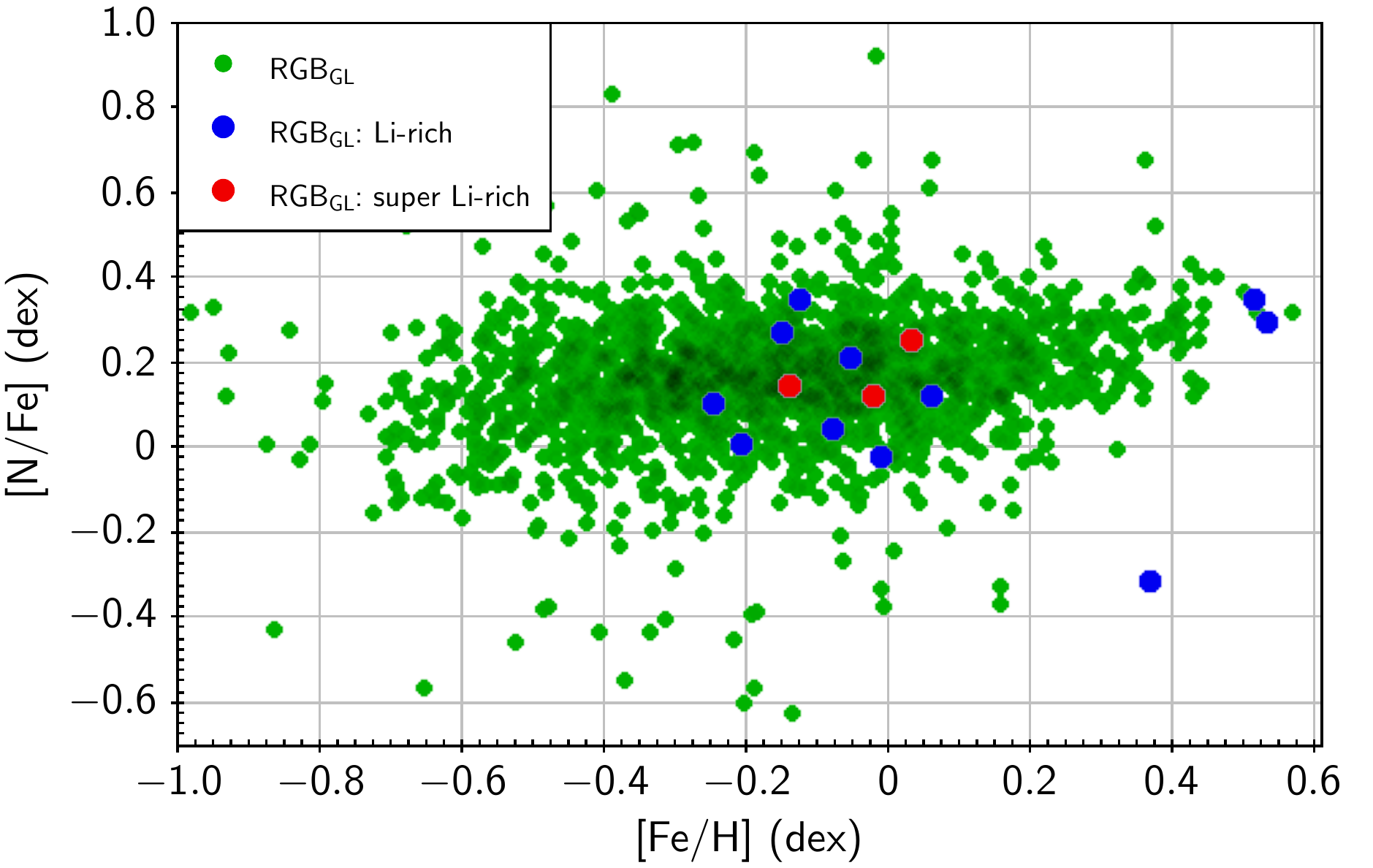}
\caption{Distribution of [N/Fe] versus [Fe/H] for sample of common giants among the {\it GALAH} and {\it LAMOST} surveys (sample RGB$_{GL}$).
\label{fig:f13}}
\end{figure}

\begin{figure*}
\includegraphics[width=1\textwidth]{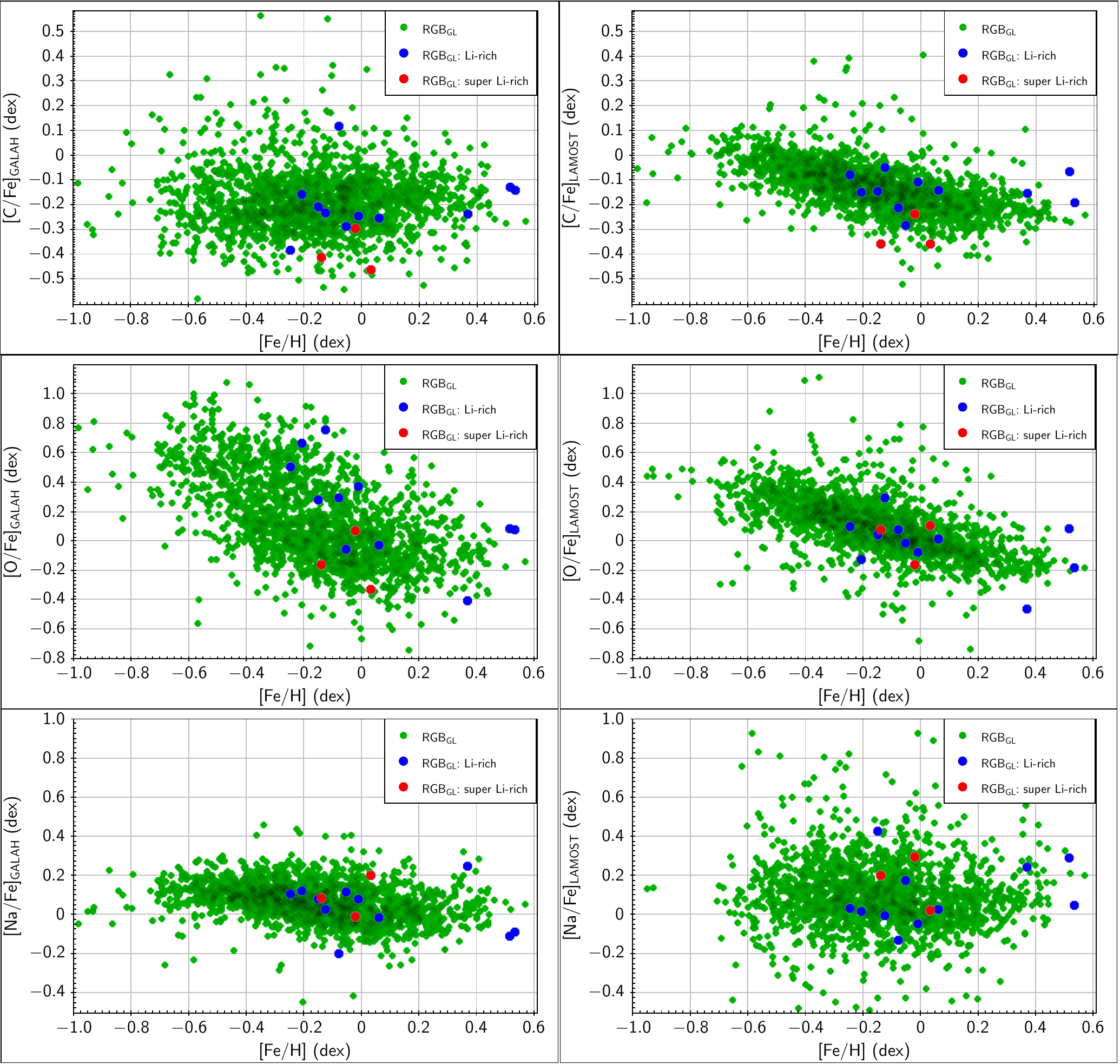}
\caption{[C/Fe], [O/Fe], and [Na/Fe] versus [Fe/H] for common giants among {\it GALAH} and {\it LAMOST}. In the left and right panels shows the abundance from the {\it GALAH} and {\it LAMOST} surveys for same sample of common giants among the {\it GALAH} and {\it LAMOST} surveys.
\label{fig:f14}}
\end{figure*}

The recent study by \cite{XiangTingRix2019arXiv190809727X} of {\it LAMOST} spectra provides  a catalogue of  abundances of 16 elements (C, N, O, Na, Mg, Al, Si, Ca, Ti, Cr, Mn, Fe, Co, Ni, Cu, and Ba, of which N is of our main interest) of about 6 million stars from $\sim$ 8 million low-resolution (R $\sim$ 1800) spectra from the {\it LAMOST} DR5. Abundances were extracted from the {\it LAMOST} spectra using training sets of high-resolution spectra provided by {\it GALAH} and {\it APOGEE} abundances. Nitrogen abundances for our giants in the {\it GALAH} survey  are provided by the {\it LAMOST}-{\it GALAH} calibrated abundances. (There is very little overlap between the {\it APOGEE} sample primarily from the northern hemisphere and the {\it GALAH} sample primarily from the southern hemisphere.) {\it LAMOST} N abundances are provided from the CH and CN bands in the blue. We cross-matched our red giant sample with the {\it LAMOST} catalogue based on a star's positional coordinates with a maximum error of one arc-sec in star's position and found about 1800 common giants. We further restricted our sample to only those giants for which estimated N abundances are available (i.e., excluded the stars with default value of $-999$ which was assigned when the code is enable to estimate the abundance, see \cite{XiangTingRix2019arXiv190809727X} for more details). This resulted in a sample of 1793 giants (here, the  RGB$\rm_{GL}$ sample) of which 14 are Li-rich (with A(Li) $\geq$ 1.8 dex) and three (with A(Li) $\geq$ 3.2 dex) are super Li-rich (with A(Li) $\geq$ 3.2 dex).

In Figure \ref{fig:f13}, N abundances provided by \cite{XiangTingRix2019arXiv190809727X} for giants common to the {\it GALAH} and {\it LAMOST} surveys are shown as a function of [Fe/H]. This figure in which giants are not separated into the luminosity classes LB$^-$ to RC$^ +$ suggests that Li-rich and super Li-rich giants share the same N abundance with the normal giants. This result is somewhat surprising given that the super Li-rich giants appear to be more C-depleted than other giants. The mean N abundance is very similar to that provided by our search of the literature (Figure \ref{fig:f11}).

Given that  the {\it GALAH} abundances are used as a training set for the low resolution {\it LAMOST} survey, one expects the [X/Fe] versus [Fe/H] relations from {\it GALAH} to be a close match to the calibrated {\it LAMOST} abundances. (In a few cases, as noted in Section \ref{sec:analysis} the {\it GALAH} abundances appear to be subject to a systematic error.) In Figure \ref{fig:f14}, we compare [X/Fe] versus [Fe/H]  for X = C, O and Na for common giants  from the two surveys, again the giants are not broken into luminosity classes.

\section{Creation of a Lithium enriched giant}\label{CreationofLiRichGiants}

For several decades, three potential reservoirs of Li that a normal giant may call upon to effect its conversion to a Li-rich giant, even a super Li-rich giant, have been proposed. A primary unresolved issue is how a reservoir may be  tapped to convert a normal star to a Li-rich giant.\\
 :-- One  reservoir is material orbiting a star -- planets and substellar masses M $\leq0.065$M$_\odot$  -- which a giant may engulf. (Brown dwarfs of mass greater than 0.065M$\odot$ destroy Li internally.) Since this material is expected to have retained its original Li (both $^7$Li and $^6$Li and also $^9$Be), capture by the giant will raise the Li (and Be) abundance of the giant's atmosphere and convective envelope. The Li abundance of the Li-rich giant depends in part on the mass of circumstellar material and its Li abundance.\\
:--  A second potential reservoir is the star's internal supply of $^3$He which, under the right circumstances, may be converted to $^7$Li by the reaction $^3$He($\alpha,\nu)^7$Be and subsequent electron-capture $^7$Be(e$^-,\nu)^7$Li. A key to successful appearance of $^7$Li at the giant's surface is that the $^7$Be synthesis by $\alpha$-captures must occur in a convective envelope which allows conversion of some $^3$He to $^7$Be but permits an adequate amount of $^7$Be (and $^7$Li) to be convected to lower temperatures and, thus, avoid near instantaneous destruction by $\alpha$-captures. Of course, the convective envelope will return $^7$Li to the deeper hotter layers and so expose it to destruction by $\alpha$-captures as additional $^3$He may be converted to $^7$Be. \\
:- The third possible reservoir of Li is a companion and more massive star which evolves to become an AGB giant. Luminous AGB stars are predicted and known to have Li enriched envelopes \citep{SmithLambert1989}. The origin of the freshly synthesized Li is again the star's $^3$He.  Then, a measure of mass transfer from the luminous AGB star to its companion could result in a Li-rich giant (or main sequence) star. Since luminous AGB stars have experienced considerable enrichment with products of the $s$-process, the Li-rich giant would be expected to show marked $s$-process  as well as Li-enrichment. None of the Li-rich giants in our {\it GALAH} survey are $s$-process enriched and, thus, we do not discuss this possible reservoir further. This binary scenario is the process by which classical Ba and CH giants are created.

\subsection{Thoughts on the external reservoir}

An extensive literature on theoretical and observational aspects on Li enrichment resulting from accretion of circumstellar material now exists with its origin traceable to Alexander's (\citeyear{Alexander1967Obs....87..238A}) suggestion that ingestion of a planet by a red giant would boost the star's Li  abundance. Theoretical studies \citep[e.g.,][]{AguileraGomez2016ApJ...829..127A} have considered ingestion of brown dwarfs, Jupiter-like and Earth-like planets. Brown dwarfs with a mass M$ \leq 0.065$M$_\odot$ are expected to preserve their Li and at least partially offset the reduction of the Li abundance arising from the first dredge-up. In some calculations, the brown dwarf is considered to be metal-rich relative to the companion star. In many calculations, Jovian planets are assumed to have the same composition as the initial main sequence star, i.e., the same mass fraction of Li. Calculations incorporating ingestion of terrestrial planets clearly adopt a composition different from that of the main sequence including a composition representative of the present Earth; the Li abundance is raised on ingestion but relative abundances of volatiles and non-volatiles may be altered. In each case, not only is the giant's surface Li abundance raised by  ingestion but the isotopic ratio $^6$Li/$^7$Li is presumably raised close to its original value and the stellar Be abundance too is raised to partially correct for the dilution  of Be by the first dredge-up.

For estimates of the Li abundance of red giants post-ingestion of circumstellar material, we turn to \cite{AguileraGomez2016ApJ...829..127A} who examined accretion of brown dwarfs and planets Jupiter and Earth by low mass stars with [Fe/H] from 0 to $-2.0$. A maximum mass to accreted material and, hence, a maximum Li abundance for the post-accretion giant is set by their demonstration that brown dwarfs with masses greater than 15 Jovian masses (15M$_J$) are ingested but dissolve in the giant's radiative core below its convective envelope.
In these calculations, a giant's main sequence progenitor is assumed to have not destroyed Li in its outer envelope. Li depletion in main sequence stars is a common phenomenon and, thus, the calculations may overestimate the Li abundance following an accretion event. Destruction of accreted Li by the giant is included in the calculations.
A 1.5M$_\odot$ giant with [Fe/H] = 0 with an initial abundance A(Li) = 3.3, the meteoritic value,  after ingestion of a 15M$_J$ brown dwarf with an initial composition 2.5 times solar (including Li) has a predicted Li abundance A(Li) = 1.8. This Li abundance is reduced to A(Li) = 1.6 if the brown dwarf has a solar composition.  The Li abundance is slightly higher for a 1M$_\odot$ giant. These abundances are insensitive to the initial stellar abundance of Li  but not to the assumed Li abundance of the brown dwarf. Of course, the giant's predicted post-accretion Li abundance is much lower after accreting a Jovian or an Earth-like planet. For a fixed initial Li abundance for the accreted material, the giant's final Li abundance is insensitive to its initial [Fe/H] and insensitive to the point on the giant branch at which ingestion begins. Indeed, the final Li abundance for at 1M$_\odot$ for a fixed initial Li abundance increases by about 0.4 dex as [Fe/H] drops from 0.0 to $-2.0$. Since the initial Li abundance decreases with decreasing [Fe/H] in the Galactic disk \citep{Lambert&Reddy2004}, the predicted Li abundance of a giant is in practice sure to decrease with decreasing [Fe/H].

These calculations, as \cite{AguileraGomez2016ApJ...829..127A} emphasize, cannot account for the most Li enriched giants and, indeed, even our lower limit A(Li) = 1.8 for a Li-rich giant is at the limit of their calculations. Of course, the giant's final Li abundance may be additionally increased by ingestion of more than a single brown dwarf, Jovian or terrestrial planet. For example, Melo et al.'s (\citeyear{MeloLaverny2005A&A...439..227M}) simpler recipe suggests that A(Li) = 2.5 is attainable with ingestion of 100 Jupiters. A couple of additional observations. First, giants with planets and including Li-rich planets with planets are known \citep{AdamowNiedzielski2018A&A...613A..47A} and the likelihood that some giants have experienced ingestion seems high and the question is how does one recognize such giants even if the Li abundance does not cross the threshold to be classified as Li-rich.

One test involves the Be abundance derived from the Be\,{\sc ii} 3131 \AA\ resonance lines. Surface Be abundance of a giant is diluted by the first dredge-up and replenished by ingestion of circumstellar material. The point that Be has not increased above first dredge-up values in Li enriched stars has been made previously \citep{MeloLaverny2005A&A...439..227M, TakedaTajitsu2017PASJ...69...74T}. This is certainly a valid point against the external reservoir as a source of Li for Li-rich and especially for super Li-rich giants. But since Be abundance determinations are relatively rare, a giant which has ingested circumstellar material including its own planets may await discovery among nearby giants.

Ingestion of Earth-like planets will change the relative abundances of the elements. A recent interpretation of the composition differences between HD 240429 and HD 240430, solar-type stars forming a comoving pair has proposed that HD 240430 has accreted 15 Earth masses of material with the composition of the Earth \citep{OhPrice-WhelanBrewer2018ApJ...854..138O}. This accretion results in HD 240430 (relative to HD 240429) having a higher abundance by about 0.2 dex of refractory elements with a condensation temperature above about 1200 K but a similar abundance of less volatile elements like C, N, O and Na. Adopting the bulk composition of the Earth from \cite{McDonough2003TrGeo...2..547M}, a good fit is obtained to the abundance differences between the two stars. Table \ref{table:t1a} and Figure \ref{fig:f2a} show that abundance differences between normal and Li enriched giants are small across elements with different condensation temperatures and absence of a correlation of abundance differences with condensation temperature provides an opportunity to set a limit on accretion of Earth-like material by the Li enriched giants. A giant's convective envelope includes about 60\% of the mass of the giant's mass \citep{AguileraGomez2016ApJ...829..127A}. Addition of 10 Earth-like planets would increase the abundance of refectory elements by about 0.05 dex relative to less volatile elements such as O and Na. Abundance differences [X/Fe] in Table 1 are generally less than 0.05 dex. For example, abundance differences for the RC sample in the [Fe/H] interval $+0.1$ to $+0.3$ dex run from $-0.03$ to $+0.04$ dex excluding C.  Thus, Li enriched giants have ingested little to no Earth-like planets.

\subsection{Thoughts on the internal reservoir}

Application of a giant's internal $^3$He supply to the puzzle of Li enriched stars demands a trigger for mixing between deep layers where conversion of $^3$He to $^7$Be occurs and the surface where an overabundance of $^7$Li (and possibly $^7$Be) occurs. As noted above, a novel result of this analysis is that many Li enriched giants share the slight C deficiency of normal giants, which is the expected consequence of the first dredge-up. It is also the case that the $^{12}$C/$^{13}$C ratio of Li-rich and super Li-rich giants span the range from 5 to 25 also exhibited by normal giants \citep{KumarReddyLambert2011, AdamowNiedzielski2018A&A...613A..47A}. This shared pattern of C elemental and isotopic abundances suggests that the $^3$He-to-$^7$Li conversion may operate independently of the processes providing the internal composition changes to C, N and O which are revealed as surface composition changes by the first dredge-up. This independence appears broken in the case of the super Li-rich giants where additional C deficiency was found but without a clear difference in the $^{12}$C/$^{13}$C ratio (Figure \ref{fig:f12}).

All $^7$Li produced from the internal reservoir comes from decay of $^7$Be with its 53 day half-life against electron capture. \cite{TakedaTajitsu2017PASJ...69...74T} noted that the isotopic wavelength shift for the Be\,{\sc ii} 3130-3131 \AA\ resonance lines suffices in a high resolution spectrum to put the $^7$Be line to the red and clear of the stable $^9$Be line.  Their two comparisons of the spectrum of  a  Li-rich giant with the spectrum of a normal giant with nearly identical atmospheric parameters showed no evidence of a $^7$Be contribution to the spectrum of the Li-rich giant but a marked but  expected $^9$Be deficiency in both stars. There may be a tantalizing hint of $^7$Be in the spectrum of two super Li-rich giants whose spectra are shown by Melo et al. (\citeyear[][compare their Figures 3 and 4]{MeloLaverny2005A&A...439..227M}). Spectra of these giants around 3130 \AA\ are rich in atomic and molecular (OH) lines and a careful accounting of these blending lines is required before claiming detection of $^7$Be or even $^9$Be.

It would be of interest to calculate from a model of a giant the probability that an observable amount of $^7$Be may be convected to the surface before it undergoes electron capture and conversion to $^7$Li.  In the convective envelope,  Be is fully ionized and thus the $^7$Be's lifetime is greatly enhanced over its natural 53 days for the neutral atom.  Transport over the stellar envelope, a distance of about 50 solar radii, in just 53 days requires a velocity of about 8 km s$^{-1}$, which would appear to be a stiff requirement but the increase in lifetime of the bare $^7$Be nucleus may reduce the demanded convective velocity to a physically relevant value.  A maximum $^7$Be abundance equal to the maximum Li abundance of A(Li) = 4.5 observed in a Li-rich K giant after six half-lives at even the natural 53  day half-life provides a $^7$Be abundance A($^7$Be) = 1.6 at the stellar surface, an abundance slightly exceeding the initial (i.e., solar) of $^9$Be. 
Thus, it may appear possible that super Li-rich giants may show detectable amounts of $^7$Be but this possibility demands that the giant be currently experiencing conversion of $^3$He to $^7$Be in its interior. On cessation of $^7$Be  production, $^7$Be in the atmosphere decays with its 53 day half life, a fleeting second with respect to the 100-200 million year life of the He-core burning clump giant. Acquisition of high-resolution spectra around 3130 \AA\ of Li enriched giants is of interest. Positive identification of $^7$Be would be proof not only that the Li production originated with the $^3$He internal reservoir but that $^3$He conversion to $^7$Be and $^3$He is occurring presently in that giant.
Ignition of $^3$He as a source of $^7$Li can not synthesis $^9$Be, the sole stable isotope of Be.

Even with the attribution of Li production to the $^3$He internal reservoir, there remains the challenge of identifying why only 1\% of giants  experience substantial Li enrichment at their surfaces. This challenge is sharpened if, as evidence is now trending, the key event in a Li-rich giant's life is its He-core flash at the tip of the red giant branch.
\cite{Casey2019ApJ...880..125C} argue that it is tidal spin-up by a stellar companion  that triggers conversion of $^3$He to $^7$Li. A merit of this proposal is that it is open to a straightforward test: radial velocity observations should reveal the companion star.

\section{Concluding remarks}\label{sec:conclusion}

In conclusion, the compositions of Li-rich and  the super Li-rich giants across the elements considered by the {\it GALAH} survey are identical but for slightly lower [C/Fe] values in the super Li-rich giants. Distribution of the [C/Fe] values for the Li-rich giants resembles that for the normal giants. This analysis of {\it GALAH} abundances appears to be the first extensive differential survey to examine compositions of Li-rich and super Li-rich giants along with normal giants in the search of differences between the Li enriched and normal giants. Although the literature on the compositions of Li enriched giants is now quite extensive, few studies have tackled a fair sample of the Li enriched giants and a comparable sample of normal giants in order to minimize systematic errors in the analyses. Among published studies, we note that by \cite{TakedaTajitsu2017PASJ...69...74T} whose sample contained 20 Li enriched giants including five super Li-rich and all but two giants with a Li abundance A(Li) $\geq 2.0$. \cite{TakedaTajitsu2017PASJ...69...74T} obtained C, O. Na, S and Zn abundances by line selections and methods previously used for a large sample of normal giants.  (These authors also derived Be abundances for five Li enriched giants  and estimated the $^6$Li/$^7$Li ratio.) Elemental abundances for  the Li-rich giants were not all exactly in line with those of normal giants.  Of  interest to understanding  the origins of Li-rich giants, the [C/Fe] values of about half of the sample were much in excess of the values for normal giants of similar metallicity. About eight Li-rich giants had a positive [C/Fe] value at metallicities for which normal giants had a negative [C/Fe] value comparable to the {\it GALAH} values. (The C abundance was provided by the C\,{\sc i} 5380 \AA\ line.)  This divergence in [C/Fe] values with the {\it GALAH} values is not understood. In principle, Li-rich giants may be created through mass transfer from a Li-rich AGB star which could be C-rich but would also be expected to be $s$-process enriched. Our sample of Li-rich giants is not $s$-process enriched. The {\it Gaia-ESO} survey of Li-rich giants \citep{CaseyRuchtiMasseronRandich2016,SmiljanicFranciosini2018} has provided information on the compositions of a few giants indicating essentially normal compositions relative to normal giants.

Our observation that the Li-rich giants and normal giants have the same compositions including for species (e.g., C, N and O) whose surface abundances are  affected by the first dredge-up suggests that the addition of Li to create a Li-rich giant occurs independently of the abundance changes wrought by CN-cycling in the deep interior. Although a few Li-rich giants may result from accretion of external material, the majority of these giants have Li abundances indicating internal production of Li from $^3$He conversion to $^7$Be and the latter's decay to $^7$Li (i.e., the Cameron-Fowler mechanism).  Demonstration that the super Li-rich giants are more severely depleted in C than the normal Li-rich giants may indicate that their production of $^7$Be was accompanied by additional CN-cycling.

We also found that the probability of becoming a Li-rich giant is independent of star's mass, although majority of the Li-rich giants are found to be low mass ($M \leq$ 2 M$_\odot$).
Both the Li-enriched and normal giants are also found to have similar projected rotational velocity which suggest that Li-enrichment in giants is not linked to scenarios such as mergers, tidal interaction between binary stars, etc. The frequency of occurrence of Li-rich giants among normal giants is about 1 per cent and, as found by \cite{Casey2019ApJ...880..125C}, this frequency increases with increasing metallicity.

As shown in Figure \ref{fig:f1}, the great majority of Li-rich giants are He-core burning giants but there are minorities with apparent connection to the red giant branch, possibly to the luminosity bump, and to the red giant branch at luminosities above the bump and to evolution along the early-AGB following He-core burning at the clump. All but one of the super Li-rich giants appears as a He-core burning or clump giant. (Low luminosity Li-rich stars in Figure \ref{fig:f1} at effective temperatures of about 5200 K may be stars for which the first dredge-up is as yet incomplete.)  Distribution of Li-rich giants in their version of Figure \ref{fig:f1} led \cite{Casey2019ApJ...880..125C} to conclude that production of a Li-rich giant does not occur exclusively at the luminosity bump  or the tip of the red giant branch, but their Li abundance estimates are from the low resolution ($\sim$ 1800) {\it LAMOST} spectra.
They stars become Li-rich  either ``at a random time on the giant branch or at the start of the helium-burning phase, and remain lithium-rich for about 2 x $10^6$ yr''. Between the tip of the red giant branch  and the He-core burning phase is the He-core flash in low mass giants. This core flash is, among some observers, a prime source of internal disturbance resulting in Li-enrichment at a giant's surface. Theoretical evidence in support of the suspicion is lacking, however. \cite{Casey2019ApJ...880..125C} argue with theoretical backing that the conversion of $^3$He to $^7$Be with eventual decay to $^7$Li is triggered by tidal interactions between the giant and a stellar companion. As \cite{Casey2019ApJ...880..125C} note in a concluding paragraph, their proposal about tidal interactions is testable by a search for binary companions to Li-rich giants. It will be interesting to see if Li-rich giants do indeed have binary companions and if the nature of the binaries accounts for the distribution of Li abundances and, in particular, for the occurrence of super Li-rich giants with their additional depletion of carbon. This will not be last time that a lithium puzzle has sent observers back to their telescopes.

\section*{Acknowledgments} \label{sec:acknowledgments}
We thank the referee for a series of helpful and constructive remarks.
Deepak would like to thank A. B. S. Reddy for helpful comments.
This work has made use of the {\it GALAH} survey which includes data acquired through the Australian Astronomical Observatory, under programmes: A/2013B/13 (The {\it GALAH} pilot survey); A/2014A/25, A/2015A/19, A2017A/18 (The {\it GALAH} survey). We acknowledge the traditional owners of the land on which the AAT stands, the Gamilaraay people, and pay our respects to elders past and present.

This work has also made use of data from the European Space Agency (ESA) mission {\it Gaia} (\url{https://www.cosmos.esa.int/gaia}), processed by the {\it Gaia} Data Processing and Analysis Consortium (DPAC, \url{https://www.cosmos.esa.int/web/gaia/dpac/consortium}). Funding for the DPAC has been provided by national institutions, in particular the institutions participating in the {\it Gaia} Multilateral Agreement.

This work has also made use of data from the {\it LAMOST} survey. The data for the {\it LAMOST} survey is acquired through the Guoshoujing Telescope. Guoshoujing Telescope (the Large Sky Area Multi-Object Fiber Spectroscopic Telescope; {\it LAMOST}) is a National Major Scientific Project built by the Chinese Academy of Sciences. Funding for the project has been provided by the National Development and Reform Commission. {\it LAMOST} is operated and managed by the National Astronomical Observatories, Chinese Academy of Sciences.

This research has made use of NASA's Astrophysics Data System.

\section*{ORCID iDs}
Lambert, D. L.: \url{https://orcid.org/0000-0003-1814-3379}\\
Reddy, B. E.: \url{https://orcid.org/0000-0001-9246-9743}\\
Deepak: \url{https://orcid.org/0000-0003-2048-9870}

\bibliographystyle{mnras}
\bibliography{ref} 




\bsp	
\label{lastpage}
\end{document}